\pdfoutput=1
\documentclass[11pt]{article}
\usepackage[final]{acl}
\usepackage{times}
\usepackage{latexsym}
\usepackage{hyperref}
\usepackage{subfig}
\usepackage[T1]{fontenc}
\newcommand\blfootnote[1]{%
  \begingroup
  \renewcommand\thefootnote{}%
  \footnotetext{#1}%
  \addtocounter{footnote}{-1}%
  \endgroup
}

\usepackage[utf8]{inputenc}

\usepackage{microtype}
\usepackage{verbatim}
\usepackage{inconsolata}

\definecolor{emerald5}{RGB}{236,247,238}   
\definecolor{emerald10}{RGB}{200,230,210}  
\definecolor{emerald15}{RGB}{164,214,185}

\usepackage{microtype}
\definecolor{lightred}{rgb}{1.0, 0.8, 0.8}
\definecolor{lightblue}{rgb}{0.8, 0.9, 1.0}
\definecolor{lightgreen}{rgb}{0.8, 1.0, 0.8}
\definecolor{lightyellow}{rgb}{1.0, 1.0, 0.8}
\definecolor{lightpurple}{rgb}{0.9, 0.8, 1.0}
\definecolor{lightorange}{rgb}{1.0, 0.9, 0.8}
\usepackage{inconsolata}

\usepackage{amsmath, amssymb, graphicx, xcolor, multirow, multicol, comment, subcaption, algorithm2e, url, float, etoolbox, adjustbox, pgf, soul, geometry, colortbl, booktabs}

\usepackage{tcolorbox}

\title{\textit{HCFD:} A Benchmark for Audio Deepfake Detection in Healthcare}

\author{
\textbf{Mohd Mujtaba Akhtar\textsuperscript{1}\thanks{Equal contribution as a first author.}} \quad
\textbf{Girish\textsuperscript{2}\footnotemark[1]} \quad
\textbf{Muskaan Singh\textsuperscript{3}\thanks{Corresponding author.}}\\
\textsuperscript{1}Veer Bahadur Singh Purvanchal University, India \quad
\textsuperscript{2}UPES, India \quad
\textsuperscript{3}Ulster University, UK\\
\texttt{\{mmakhtar.research, girish.research.pr\}@gmail.com}, \quad \texttt{m.singh@ulster.ac.uk}
}

\begin{document}
\maketitle
\begin{abstract}
In this study, we present Healthcare CodecFake Detection (HCFD), a new task for detecting codec-fakes under pathological speech conditions. We intentionally focus on codec-based synthetic speech in this work, since neural codec decoding forms a core building block in modern speech generation pipelines. First, we release Healthcare CodecFake, the first pathology-aware dataset containing paired real and NAC-synthesized speech across multiple clinical conditions and codec families. Our evaluations show that SOTA codec-fake detectors trained primarily on healthy speech perform poorly on Healthcare CodecFake, highlighting the need for HCFD-specific models. Second, we demonstrate that PaSST outperforms existing speech-based models for HCFD, benefiting from its patch-based spectro-temporal representation. Finally, we propose \textbf{\texttt{PHOENIX-Mamba}}, a geometry-aware framework that models codec-fakes as multiple self-discovered modes in hyperbolic space and achieves the strongest performance on HCFD across clinical conditions and codecs. Experiments on HCFK show that \textbf{\texttt{PHOENIX-Mamba}} (PaSST) achieves the best overall performance, reaching 97.04 Acc on E-Dep, 96.73 on E-Alz, and 96.57 on E-Dys, while maintaining strong results on Chinese with 94.41 (Dep), 94.40 (Alz), and 93.20 (Dys). This geometry-aware formulation enables self-discovered clustering of heterogeneous codec-fake modes in hyperbolic space, facilitating robust discrimination under pathological speech variability. \textbf{\texttt{PHOENIX-Mamba}} achieves topmost performance on the HCFD task across clinical conditions and codecs.

\end{abstract}

\section{Introduction}

Imagine a clinical voice sample collected to track disease progression being replaced by a codec-generated replica that sounds natural to clinicians and passes automated checks. This scenario is increasingly plausible as neural audio codecs and modern speech generation pipelines enable high-fidelity synthetic speech at scale \cite{ZeghidourLOST22,DefossezCSA23,BorsosMVKPSRTGTZ23,wang2023neural}. While audio deepfake detection has progressed rapidly, its robustness under pathology-driven speech variability—a defining property of healthcare audio—remains underexplored. Pathological speech systematically alters prosody and articulation, which can obscure codec-induced cues and cause detectors trained on healthy speech to fail in clinical settings. Recent benchmarks further reveal robustness gaps under realistic conditions, including telephony-based deepfakes \cite{yamagishi21_asvspoof} and in-the-wild conversational speech \cite{wang24_asvspoof}, which reflect realistic recording conditions and spontaneous speech patterns. Moving to clinical speech, a key barrier is the lack of pathology-aware healthcare deepfake data. While recent work has advanced codec-fake detection, the datasets used for training and evaluation are largely built on healthy, clean speech, with little pathology-aware healthcare deepfake data. As a result, detectors are rarely exposed to disease-related acoustic variation, even though pathological speech differs markedly in prosody, articulation, and phonation. These shifts can change how codec traces appear, limiting transfer to clinical settings and leaving healthcare speech under-served in the current codec-fake detection ecosystem. Healthcare speech is increasingly captured beyond controlled clinic visits—via telehealth consultations, contact-center triage, and remote screening—then transmitted and stored through real-world audio pipelines. In such settings, voice serves a dual role: it is both a clinical signal for monitoring neurocognitive and motor disorders and, in many deployments, an identity signal for patient or staff verification. This combination makes healthcare audio a practical target for modern voice-based attacks. Practitioner and industry analyses warn that rapidly advancing AI voice synthesis and cloning can enable social-engineering and account-takeover attempts against healthcare services, underscoring the need for pathology-aware safeguards that remain reliable under clinically realistic variability \footnote{Relevant practitioner/industry discussions:
\href{https://www.pindrop.com/article/voice-biometric-authentication-patient-privacy/}{Pindrop},
\href{https://hitconsultant.net/2024/08/08/ai-voice-deepfakes-waging-war-on-healthcare/}{HIT Consultant},
\href{https://voicebiometrics.ai/blogs/voice-biometrics-in-health-care/}{VoiceBiometrics.ai},
\href{https://www.openaccessgovernment.org/the-critical-role-of-voice-biometrics-in-healthcare-security/189376/}{OpenAccessGovernment}.}. Motivated by this real-world risk surface, we study whether current codec-fake detectors remain reliable when speech is shaped by clinically realistic variability, and we introduce a pathology-aware benchmark and framework to address this gap. \par
As a remedy for this gap, in this work, we present Healthcare CodecFake (HCFK) and define Healthcare CodecFake Detection (HCFD) as a task for codec-fake detection under pathological speech conditions. Healthcare CodecFake provides paired bona fide and Neural Audio Codecs (NACs)-synthesized speech across multiple clinical conditions and codec families. We benchmark state-of-the-art codec-fake (CF) detectors, including PaSST \citet{koutini22_interspeech}, on HCFK and observe a pronounced performance drop, revealing that current CF detectors are not designed for pathology-driven clinical audio. Healthcare speech samples inherently reflect condition-dependent changes in prosody, articulation, and phonation (e.g., altered voice quality, reduced intelligibility, and atypical temporal patterns), which can distort or conceal the subtle traces introduced by NACs, thereby highlighting the limitations of existing CF detection approaches and underscoring the necessity for pathology-aware, healthcare-specific CF detection frameworks. Next, \textit{we hypothesize that the use of pretrained audio representations mitigates the impact of pathology-driven variability by separating condition-specific speech characteristics from codec-related cues, resulting in improved generalization across clinical conditions.} To validate this hypothesis, we conduct a comprehensive comparative study across a diverse set of pretrained audio models, including both speech-focused and audio–language (multimodal) PTMs, under the HCFD setting. Through extensive experiments on HCFK spanning multiple clinical conditions and codec families, we show that PTM-based representations offer more reliable performance than conventional CF detectors, supporting our hypothesis. Despite the gains offered by PTMs, residual performance gaps under clinical variability indicate that representation choice alone is insufficient, prompting us to develop a dedicated pathology-aware detection framework. To this end, we propose \textbf{\texttt{PHOENIX-Mamba:}} \textbf{P}rototypical \textbf{H}yperbolic \textbf{O}rganization for \textbf{E}vidence \textbf{N}ormalization and \textbf{I}nference using e\textbf{X}ponential-map, a framework tailored to the challenges of codec-fake detection in pathological speech. \textbf{\texttt{PHOENIX-Mamba}} integrates long-context temporal modeling with geometry-aware, prototype-based clustering to capture the heterogeneous structure of codec artifacts in clinical speech. By organizing learned evidence representations into multiple fake modes within a hyperbolic space, the framework separates codec-induced cues from disease-related acoustic effects in a principled manner. This design enables consistent inference across clinical conditions and codec families, addressing failure modes that arise in pathology-agnostic detection pipelines. Extensive experiments on HCFK demonstrate that \textbf{\texttt{PHOENIX-Mamba}} consistently delivers stronger performance than state-of-the-art codec-fake detectors and strong PTM baselines on the HCFD task.
\noindent \textbf{Our Key contributions are as follows:} \newline
\noindent (i) We present Healthcare CodecFake (HCFK) and formally define the novel task of Healthcare CodecFake Detection (HCFD), addressing a critical gap in pathology-aware healthcare deepfake detection.; (ii) We benchmark state-of-the-art codec-fake (CF) detectors on HCFK and demonstrate substantial performance degradation, highlighting the need for pathology-aware, healthcare-specific detection frameworks.; (iii) We hypothesize and validate that pretrained audio representations (speech-focused and audio–language PTMs), due to their large-scale pretraining, are better suited for HCFD; a comprehensive comparison across diverse PTMs supports this claim.; (iv) We propose \textbf{\texttt{PHOENIX-Mamba}}—a pathology-aware detection framework that integrates long-context temporal modeling with hyperbolic, prototype-based clustering to capture heterogeneous codec artifacts, achieving consistently stronger performance than state-of-the-art CF detectors and strong PTM baselines on HCFD.

\blfootnote{Dataset access, code, and evaluation resources are provided at \url{https://helixometry.github.io/HCFD/}.}

\section{Related Work}
    \citet{wu24p_interspeech} and \citet{lu24f_interspeech} showed that vocoder-trained deepfake detectors generalize poorly to codec-synthesized speech. \citet{wu24p_interspeech} introduced a VCTK-based codec-synthesized benchmark evaluated with AASIST, while \citet{lu24f_interspeech} extended benchmark construction to VCTK and AISHELL3 and compared AASIST and LCNN using mel-spectrogram and Wav2vec2 features. 
Subsequent studies extended CodecFake benchmarks by increasing codec diversity \citep{chen2025codecfake+} and explored unified semantic--acoustic representations for detection, e.g., via SASTNet \citep{chen2025towards}. 
\citet{xie2025codecfake} considered unified detection across codec- and vocoder-synthesized speech, reporting improved cross-mechanism generalization with sharpness-aware optimization. Empirical studies have documented that audio deepfake detectors exhibit non-trivial performance degradation when evaluated under acoustic mismatches relative to their training and validation conditions~\citep{li2025ADD}.

\section{Healthcare Codecfake Dataset}
In this section, we describe the datasets, neural audio codecs, and synthesis pipeline used to construct HCFK. All recordings are sourced from established corpora accessed under their respective data-use agreements; we collect no new data, include no personally identifiable information, and generate synthetic audio solely for research and evaluation. Our intent is defensive: to support detection and risk mitigation for malicious use of AI-generated healthcare speech; we explicitly discourage misuse of the methods or generated samples.  \newline
\noindent \textit{Access to the underlying corpora remains subject to their original licenses, consent provisions, and access agreements.} To support reproducibility, we will provide the exact split files together with the full codec-generation pipeline, including preprocessing details, codec configurations, and reconstruction steps, so that HCFK can be recreated deterministically given approved access to the source datasets.

\subsection{Healthcare Speech Datasets}
\label{HSDS}
We select benchmark healthcare speech datasets spanning multiple clinical conditions to construct a realistic evaluation setting for healthcare-oriented codec-fake detection. The benchmark includes corpora in English and Chinese, enabling cross-lingual evaluation in healthcare speech.

\noindent \textbf{Depression}: we use DAIC-WOZ \cite{gratch-etal-2014-distress} as the English corpus, consisting of 189 semi-structured interviews with the virtual interviewer Ellie. For Chinese, we use the EATD-Corpus \cite{9746569}, which contains interview-style responses from 162 volunteers with SDS-based depression annotations.

\noindent \textbf{Alzheimer:} we use ADReSS/ADReSSo \cite{luz20_interspeech} as the English corpus, a widely used benchmark for Alzheimer’s disease detection that provides standardized and balanced audio samples derived from the DementiaBank Pitt “Cookie Theft” picture-description task. For Chinese, we use NCMMSC \cite{10.1145/3726302.3730313}, a Mandarin dementia benchmark with clinically annotated recordings for cognitive impairment assessment. \par

\noindent \textbf{Dysarthria:} we use TORGO \cite{10.1007/s10579-011-9145-0} as the English corpus, which contains speech from individuals with dysarthria. For Chinese, we use the Chinese Dysarthria Speech Database (CDSD) \cite{wan24b_interspeech}, a large-scale Mandarin corpus comprising approximately 133 hours of recordings.\par
\noindent Across all conditions, the original recordings are treated as bona fide speech, and paired codec-generated samples are synthesized from the same utterances using our codec pipeline.

\subsection{Neural Audio Codecs}
Following \citet{lu24f_interspeech} and \citet{wu24p_interspeech}, we use the same family of neural audio codecs, focusing on state-of-the-art, publicly released models that are easy to reproduce and widely used. \newline
\noindent \textbf{Speechtokenizer} \cite{zhang2024speechtokenizer} \footnote{\url{https://github.com/ZhangXInFD/SpeechTokenizer.git}}: It is a unified speech tokenizer built on an RVQ-GAN style neural codec. The model uses an EnCodec-based convolutional encoder--decoder backbone with Residual Vector Quantization (RVQ). \newline
\noindent \textbf{Descript Audio Codec} \cite{kumar2024high} \footnote{\url{https://huggingface.co/descript/dac_16khz}}: It is a VQ-GAN based neural audio codec targeting high-fidelity reconstruction. The approach discretizes encoder features with RVQ and trains the generator using adversarial learning alongside multi-scale frequency-domain criteria to suppress codec artifacts. \newline
\noindent \textbf{Encodec} \cite{defossez2022high} \footnote{\url{https://huggingface.co/facebook/encodec_24khz}}: It is a streaming, high-quality neural codec that couples a convolutional encoder--decoder with RVQ discretization. Training combines time-domain and frequency-domain reconstruction criteria with a spectrogram adversary for improved perceptual quality. \newline  
\noindent \textbf{Soundstream} \cite{zeghidour2021soundstream} \footnote{\url{https://github.com/haydenshively/SoundStream}}: It is an end-to-end neural codec tailored for low-bitrate speech compression. The model combines an encoder--decoder backbone with Residual Vector Quantization (RVQ) and multi-scale STFT discriminators, enabling high perceptual quality under aggressive compression (3--18~kbps). \newline
\noindent \textbf{Funcodec} \cite{du2024funcodec} \footnote{\url{https://github.com/modelscope/FunCodec}}: It is an open-source neural speech codec toolkit built to make modern codec models easy to train, reproduce, and integrate into downstream pipelines. It extends the FunASR ecosystem and provides unified training recipes and inference scripts. \newline
\noindent \textbf{Audiodec} \cite{wu2023audiodec} \footnote{\url{https://github.com/facebookresearch/AudioDec}}: It is a high-quality neural codec formulated as an end-to-end autoencoder. Training proceeds in two phases: it first learns the encoder--decoder with metric-based objectives, and then applies an adversarial refinement stage that updates only the decoder.   \newline
\noindent \textbf{SNAC} \cite{siuzdak2024snac} \footnote{\url{https://huggingface.co/hubertsiuzdak/snac_44khz}}: It is a multi-scale NAC that extends standard RVQ by allowing different quantizers. Concretely, it forms a hierarchical token representation by quantizing coarse-to-fine structure at multiple frame rates.  \newline
Detailed codec configurations and checkpoints used in our experiments are summarized in Appendix~\ref{sec:neural_audio_codecs}. For reproducibility, we provide a consolidated repository of the NAC resources used in data generation: \footnote{\url{https://github.com/CodeVault-girish/Neural-Codecs.git}}.

\subsection{Health Care Codecfake Data Generation Pipeline}
We construct HCFK using a controlled resynthesis protocol, building on established CodecFake-style dataset generation practices \cite{wu24p_interspeech}. Specifically, each bona fide pathological utterance is passed through a diverse set of NACs to produce paired codec-synthesized counterparts, reflecting the codec front-ends commonly adopted in modern audio language modeling pipelines \cite{borsos2023audiolm, lu24f_interspeech}. We begin with the healthcare speech corpora described in Section~\ref{HSDS}, spanning multiple pathological conditions and two languages (English and Chinese). Each segmented utterance is treated as a bona fide reference sample. To generate codec-synthesized spoofed counterparts, we employ a codec synthesis--resynthesis loop: for each waveform, we first pass the signal through the pre-trained encoder of a NAC to obtain a discrete latent representation, and then decode it back to the waveform domain using the corresponding decoder. We then decode the discrete representation using the corresponding codec decoder to reconstruct the signal, producing a codec-generated counterpart of the original clinical utterance. 
This reconstruction preserves most of the semantic content and speaker/subject traits, while introducing subtle artifacts arising from quantization and bandwidth constraints in the codec. As a result, HCFK contains high-fidelity yet spoofed pathological speech that reflects realistic codec-mediated generation pathways. We repeat this resynthesis procedure across a suite of NAC models, producing multiple codec-specific variants of HCFK. Concretely, for each bona fide utterance, we generate one paired codec-synthesized counterpart per codec, yielding a consistent one-to-one mapping between the source sample and each codec condition. For DAIC-WOZ \cite{gratch-etal-2014-distress}, which follows an interview-based evaluation protocol, we generate codec-synthesized samples separately within each predefined subset to keep train/dev/test strictly isolated. For ADReSS/ADReSSo \cite{luz20_interspeech} and NCMMSC \cite{10.1145/3726302.3730313}, where standard partitions are released with the benchmarks, we retain these official splits and synthesize the corresponding codec-generated audio within each split. We keep the provided splits unchanged and synthesize the codec-generated audio within each split. Across all settings, train/dev/test splits are strictly speaker-disjoint. Bona fide test utterances come from speakers unseen during training, and codec-generated samples are synthesized only from utterances within the same split. This preserves speaker disjointness for both bona fide and fake classes and ensures that evaluation reflects generalization to unseen speakers rather than speaker-specific memorization.

\begin{figure*}[!hbt]
    \centering
    \includegraphics[width=0.658\linewidth]{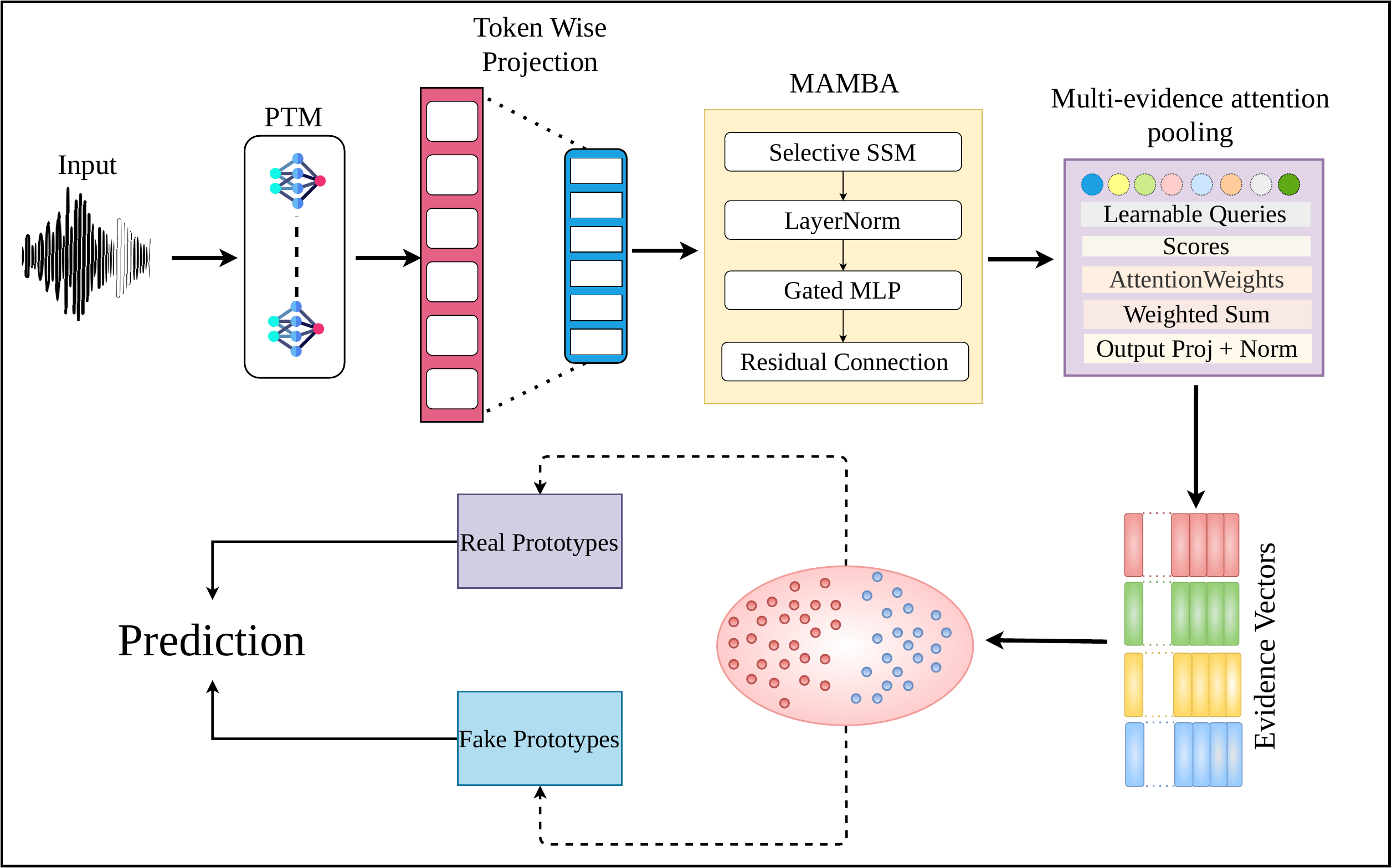}
    \caption{Proposed Framework: \textbf{PHOENIX-Mamba}}
    \label{archi}
\end{figure*}
\section{Methodology}
In this section, we provide the methodological details of our study. We begin by outlining the PTMs considered. Next, we define the downstream baselines trained on individual PTM representations. Finally, we present \textbf{\texttt{PHOENIX-Mamba}}, a geometry-aware detector that retains multiple localized evidences and models diverse fake modes using hyperbolic prototypes.

\begin{table}[!hbt]
\centering
\scriptsize
\setlength{\tabcolsep}{2.5pt}
\renewcommand{\arraystretch}{1.15}
\begin{tabular}{l|cc|cc|cc}
\toprule
\multicolumn{1}{c|}{\multirow{3}{*}{\textbf{Method}}} &
\multicolumn{2}{c|}{\textbf{Dep}} &
\multicolumn{2}{c|}{\textbf{Alz}} &
\multicolumn{2}{c}{\textbf{Dys}} \\
\cmidrule(lr){2-3}\cmidrule(lr){4-5}\cmidrule(lr){6-7}
\multicolumn{1}{c|}{} &
\textbf{Acc$\uparrow$} & \textbf{F1$\uparrow$} &
\textbf{Acc$\uparrow$} & \textbf{F1$\uparrow$} &
\textbf{Acc$\uparrow$} & \textbf{F1$\uparrow$} \\
\cmidrule(lr){2-3}\cmidrule(lr){4-5}\cmidrule(lr){6-7}
\multicolumn{1}{c}{} & \multicolumn{6}{|c}{\textbf{English}} \\
\midrule
AASIST (Tr. on CF)         & 48.62 & 44.03 & 34.19 & 32.51 & 36.71 & 34.39 \\
AASIST (Tr. in-domain)     & 60.84 & 57.92 & 52.14 & 49.93 & 56.07 & 54.49 \\
AASIST (wav2vec2.0)        & 63.55 & 51.29 & 57.76 & 54.98 & 59.35 & 57.16 \\
RawNet2                    & 60.46 & 50.27 & 58.96 & 53.87 & 57.24 & 54.68 \\
LCNN                       & 62.79 & 53.75 & 60.59 & 55.27 & 59.79 & 55.62 \\
SAMO                       & 64.37 & 54.78 & 61.89 & 58.41 & 60.27 & 56.73 \\
\midrule
\multicolumn{1}{c|}{\textbf{Method}} & \multicolumn{6}{c}{\textbf{Chinese}} \\
\midrule
AASIST (Tr. on CF)         & 45.81 & 42.38 & 30.79 & 29.23 & 33.96 & 32.54 \\
AASIST (Tr. in-domain)     & 58.06 & 54.79 & 48.51 & 46.12 & 54.84 & 51.03 \\
AASIST (wav2vec2.0)        & 60.82 & 49.31 & 53.90 & 52.29 & 56.11 & 54.62 \\
RawNet2                    & 58.33 & 47.83 & 54.51 & 49.77 & 56.71 & 52.64 \\
LCNN                       & 60.32 & 46.98 & 56.29 & 50.08 & 58.64 & 54.22 \\
SAMO                       & 62.18 & 49.82 & 59.52 & 53.79 & 61.45 & 54.78 \\
\bottomrule
\end{tabular}
\caption{Performance of prior detector baselines on three healthcare speech tasks: Depression (Dep), Alzheimer’s (Alz), and Dysarthria (Dys), evaluated separately for English and Chinese. We report Accuracy and macro-F1. ``Tr. on CF'' denotes training on the standard CodecFake benchmark and evaluation on HCFK. ``Tr. in-domain'' denotes the within-dataset setting, where for each language--condition subset the model is trained on the training split, selected on the validation split, and evaluated on the held-out test split. In addition to AASIST, we include representative waveform-based (RawNet2), spectrogram-based (LCNN), and robust one-class/generalization-oriented (SAMO) anti-spoofing baselines.}
\label{tableaasist}
\end{table}

\noindent \textbf{Pre-Trained Models}: We evaluate a diverse set of strong pre-trained encoders that have shown competitive performance across speech/audio benchmarks. Specifically, we consider self-supervised speech encoders WavLM\footnote{\url{https://huggingface.co/microsoft/wavlm-base}} \cite{chen2022wavlm} and Wav2vec~2.0\footnote{\url{https://huggingface.co/facebook/wav2vec2-base}} \cite{baevski2020wav2vec}, the Whisper model\footnote{\url{https://huggingface.co/openai/whisper-base}} \cite{radford2023robust} (using its encoder representations), a supervised speaker-embedding extractor X-vector\footnote{\url{https://huggingface.co/speechbrain/spkrec-xvect-voxceleb}} \cite{8461375}, and the spectrogram-based audio transformer PaSST\footnote{\url{https://github.com/kkoutini/PaSST}} \cite{koutini2021efficient}. All audio inputs are resampled to 16~kHz before being passed to the respective encoders. For WavLM, Wav2vec~2.0, and Whisper, we keep the PTMs frozen, extract the last hidden-state sequence (from the encoder in the case of Whisper), and obtain an utterance-level representation by average pooling over time; this pooled embedding is then fed to the downstream classifier. For X-vector, we use the VoxCeleb-trained model as a frozen feature extractor and directly feed the resulting fixed-dimensional utterance embedding to the classifier. For PaSST, we compute spectrogram inputs as required by the model and use its pooled representation as the utterance-level embedding for classification. The resulting feature vector sizes are: 768 for WavLM, Wav2vec~2.0, and PaSST; and 512 for Whisper and X-vector.

\subsection{Individual Representation Modeling} 
We establish strong baselines by training standard downstream classifiers on individual pre-trained speech representations, using a lightweight 1D-CNN (two Conv1D–BN–activation–max-pooling blocks followed by flattening and a small dense predictor) and an FCN that removes the convolutional front-end while keeping the same dense predictor. Detailed hyperparameter settings and model configurations are provided in the Appendix~\ref{sec:hyperparameters}.
\begin{table*}[!hbt]
\centering
\scriptsize
\setlength{\tabcolsep}{6pt}
\renewcommand{\arraystretch}{1.15}
\begin{tabular}{l|cc|cc|cc|cc|cc|cc}
\toprule
\multicolumn{1}{c|}{\multirow{3}{*}{\textbf{PTMs}}} &
\multicolumn{4}{c|}{\textbf{Dep}} &
\multicolumn{4}{c|}{\textbf{Alz}} &
\multicolumn{4}{c}{\textbf{Dys}} \\
\cmidrule(lr){2-5}\cmidrule(lr){6-9}\cmidrule(lr){10-13}
\multicolumn{1}{c|}{} &
\multicolumn{2}{c|}{\textbf{FCN}} & \multicolumn{2}{c|}{\textbf{CNN}} &
\multicolumn{2}{c|}{\textbf{FCN}} & \multicolumn{2}{c|}{\textbf{CNN}} &
\multicolumn{2}{c|}{\textbf{FCN}} & \multicolumn{2}{c}{\textbf{CNN}} \\
\cmidrule(lr){2-3}\cmidrule(lr){4-5}\cmidrule(lr){6-7}\cmidrule(lr){8-9}\cmidrule(lr){10-11}\cmidrule(lr){12-13}
\multicolumn{1}{c|}{} &
\textbf{Acc $\uparrow$} & \textbf{F1 $\uparrow$} & \textbf{Acc $\uparrow$} & \textbf{F1 $\uparrow$} &
\textbf{Acc $\uparrow$} & \textbf{F1 $\uparrow$} & \textbf{Acc $\uparrow$} & \textbf{F1 $\uparrow$} &
\textbf{Acc $\uparrow$} & \textbf{F1 $\uparrow$} & \textbf{Acc $\uparrow$} & \textbf{F1 $\uparrow$} \\
\midrule
\multicolumn{13}{c}{\textbf{ENGLISH}} \\
\midrule
WLM    & 69.51 & 66.83 & 71.42 & 69.13 & 57.21 & 55.62 & 59.80 & 58.17 & 63.83 & 60.47 & 65.77 & 63.21 \\
WV2 & 73.29 & 71.95 & 76.09 & 73.61 & 60.78 & 59.42 & 63.46 & 61.53 & 66.78 & 64.89 & 69.35 & 66.92 \\
WHI  & 71.58 & 68.46 & 73.38 & 71.56 & 63.84 & 59.72 & 65.98 & 62.81 & 67.15 & 65.91 & 69.48 & 68.26 \\
XVE  & 67.05 & 66.34 & 69.84 & 68.29 & 56.26 & 54.98 & 58.32 & 57.64 & 63.52 & 62.39 & 66.80 & 64.69 \\
PST    & 76.71 & 73.89 & 78.98 & 76.62 & 66.09 & 62.57 & 67.94 & 65.27 & 69.27 & 67.86 & 71.03 & 70.54 \\
\midrule
\multicolumn{13}{c}{\textbf{CHINESE}} \\
\midrule
WLM    & 66.04 & 63.79 & 69.28 & 67.59 & 55.03 & 53.81 & 56.34 & 53.67 & 60.41 & 58.03 & 63.19 & 60.47 \\
WV2 & 71.68 & 69.84 & 72.94 & 69.20 & 58.33 & 55.89 & 61.27 & 60.02 & 62.97 & 59.38 & 65.88 & 64.23 \\
WHI  & 67.93 & 64.02 & 70.51 & 66.82 & 59.91 & 56.24 & 63.53 & 61.39 & 65.34 & 64.01 & 66.41 & 64.60 \\
XVE  & 64.52 & 62.49 & 67.08 & 64.96 & 54.68 & 50.91 & 54.67 & 51.83 & 59.96 & 56.73 & 64.27 & 61.39 \\
PST    & 74.16 & 70.63 & 75.69 & 72.19 & 63.49 & 61.05 & 65.71 & 64.24 & 67.52 & 66.10 & 67.36 & 65.02 \\
\bottomrule
\end{tabular}
\caption{Performance on HCFK. Abbreviations: Dep = Depression; Alz = Alzheimer’s; Dys = Dysarthria; WLM = WavLM; WV2 = wav2vec~2.0; WHI = Whisper; XVE = x-vector; PST = PaSST. The same abbreviations are used consistently in Table~\ref{mambaphoenix}.}

\label{tab:ptm_clinical_results}
\end{table*}

\subsection{Proposed Framework: \textbf{\texttt{PHOENIX-Mamba}}}
We propose \textbf{\texttt{PHOENIX-Mamba}}, a geometry-aware approach for detecting codec-generated speech in healthcare data. The complete pipeline is depicted in Figure~\ref{archi}. Given an input utterance $x$, we first extract a sequence of latent features $X=[x_1,\dots,x_T]\in\mathbb{R}^{T\times D}$ using an upstream encoder. We then apply a token-wise alignment (adapter) map $\phi:\mathbb{R}^{D}\rightarrow\mathbb{R}^{d}$ to obtain $U=\phi(X)\in\mathbb{R}^{T\times d}$. The aligned sequence is passed through a Mamba-style selective state-space backbone $f_\theta$ to build context-enriched representations $Z=f_\theta(U)=[z_1,\dots,z_T]\in\mathbb{R}^{T\times d}$. Instead of collapsing the full sequence into a single pooled vector, \textbf{\texttt{PHOENIX-Mamba}} compresses $Z$ into a small set of $M$ evidence vectors $E=[e_1,\dots,e_M]\in\mathbb{R}^{M\times d}$. This evidence construction is implemented with a learnable pooling operator: $e_m=\sum_{t=1}^{T} a_{m,t} z_t$, where $a_{m,t}\ge 0$ and $\sum_{t=1}^{T} a_{m,t}=1$. The weights $a_{m,t}$ are produced by a differentiable scoring mechanism. This design allows the model to retain multiple localized cues that may be unevenly distributed across the utterance.
\noindent To model heterogeneity in codec artifacts, we embed each evidence vector into a metric manifold.  
\noindent We use the Poincar\'e ball $\mathcal{M}=\mathbb{B}_c^h=\{v\in\mathbb{R}^h:\; c\lVert v\rVert^2<1\}$ with curvature $-c$. Each evidence vector is mapped to the manifold using a differentiable projection $\psi:\mathbb{R}^{d}\rightarrow\mathcal{M}$, giving $h_m=\psi(e_m)\in\mathcal{M}$ and $H=[h_1,\dots,h_M]\in\mathcal{M}^M$. In practice, we project to the tangent space and apply the exponential map at the origin: $h_m=\mathrm{Exp}_0^c(W e_m)$, where $\mathrm{Exp}_0^c(y)=\tanh(\sqrt{c}\lVert y\rVert)\,\frac{y}{\sqrt{c}\lVert y\rVert}$. Distances are computed using the hyperbolic geodesic distance $d_c(\cdot,\cdot)$. We perform classification using prototype-based reasoning in $\mathcal{M}$. We parameterize a single negative prototype $p_{-}\in\mathcal{M}$ for the real class and $K$ positive prototypes $\{p_{+,1},\dots,p_{+,K}\}\subset\mathcal{M}$ to capture diverse fake modes. For each evidence point $h_m$, we compute soft responsibilities over the positive prototypes using a temperature-controlled distance softmax: 
\begin{equation}
q_{m,k}
=
\frac{\exp\!\left(-d_{\mathcal{M}}(h_m,p_{+,k})/\tau\right)}
{\sum_{j=1}^{K}\exp\!\left(-d_{\mathcal{M}}(h_m,p_{+,j})/\tau\right)}
\end{equation} \newline
\noindent This allows \textbf{\texttt{PHOENIX-Mamba}} to self-discover multiple clusters within the fake class using only binary supervision. We compute evidence-level scores using manifold distances. The negative score is defined as $s_{-}(h_m)=-d_{\mathcal{M}}(h_m,p_{-})$. The positive score is computed as a smooth soft-min over the $K$ positive modes: $s_{+}(h_m)=\log\sum_{k=1}^{K}\exp(-d_{\mathcal{M}}(h_m,p_{+,k})/\tau)$. We aggregate these scores across the $M$ evidence vectors to obtain instance-level logits $S_{-}=\frac{1}{M}\sum_{m=1}^{M} s_{-}(h_m)$ and $S_{+}=\frac{1}{M}\sum_{m=1}^{M} s_{+}(h_m)$. The final probability is computed via a softmax: $P(y=+\mid x)=\mathrm{softmax}([S_{-},S_{+}])_{+}$. \newline
\noindent We train \textbf{\texttt{PHOENIX-Mamba}} end-to-end using only real/fake labels. The base objective is the cross-entropy classification loss $\mathcal{L}_{\mathrm{cls}}$ on logits $[S_{-},S_{+}]$. To encourage compact and meaningful positive clusters, we introduce a geometry-aware clustering loss:
\begin{equation}
\begin{aligned}
\mathcal{L}_{\mathrm{cluster}}
&=
\frac{1}{M}\sum_{m=1}^{M}\sum_{k=1}^{K} q_{m,k}\, d_{\mathcal{M}}(h_m,p_{+,k}) \\
&\quad +\;
\gamma\,\frac{1}{M}\sum_{m=1}^{M}\sum_{k=1}^{K} q_{m,k}\log q_{m,k}
\end{aligned}
\end{equation}
\noindent The first term pulls evidence embeddings toward their assigned positive prototypes, while the entropy term controls assignment sharpness. To avoid prototype collapse and maintain separation between modes, we add a repulsion loss:
\begin{equation}
\begin{aligned}
\mathcal{L}_{\mathrm{sep}}
&=
\sum_{1\le i<j\le K}\exp\!\bigl(-d_{\mathcal{M}}(p_{+,i},p_{+,j})\bigr) \\
&\quad +\;
\sum_{k=1}^{K}\exp\!\bigl(-d_{\mathcal{M}}(p_{+,k},p_{-})\bigr)
\end{aligned}
\end{equation}
The total loss is given by:
$\mathcal{L}=\mathcal{L}_{\mathrm{cls}}+\lambda\,\mathcal{L}_{\mathrm{cluster}}+\beta\,\mathcal{L}_{\mathrm{sep}}$,
where $\lambda,\beta,\gamma\ge 0$ control the contribution of geometry-aware regularization. Under this objective, the backbone learns evidence representations that support robust discrimination, while the positive prototypes self-organize into multiple modes that capture heterogeneous codec artifacts. The trainable parameters for \textbf{\texttt{PHOENIX-Mamba}} range from 2 M to 5 M, depending on the input representation dimension.

\subsection{Training Details and Hyperparameters}
We train the model with AdamW for 20 epochs, using a batch size of 32, weight decay of 0.01, and gradient clipping at 1.0. For evaluation, Accuracy and macro-F1 are computed with a validation-selected decision threshold, while EER is computed from the same score distributions. All main experiments use a single shared default configuration. The final hyperparameter values are summarized in Appendix~\ref{sec:hyperparameters}.

\section{Results \& Discussion}
\label{tab:phoenix_mamba_results}

\noindent \textbf{Generalization from prior codec-deepfake detectors}: We first examine whether a detector trained under standard codec-deepfake benchmarks \cite{9747766} can generalize to pathological healthcare speech, a setting that couples codec artifacts with clinically-driven acoustic deviations. Table~\ref{tableaasist} reports results for prior detector baselines on the English and Chinese subsets across the three clinical tasks. Training AASIST on the original CodecFake distribution yields near-chance performance across all three tasks; on the English subset the scores are 48.62/34.19/36.71, while on the Chinese subset they are 45.81/30.79/33.96 for Dep/Alz/Dys, respectively. This behavior indicates a pronounced distribution shift: codec-induced cues in pathological speech are confounded by condition-specific acoustics and recording variability, limiting the reliability of models transferred from healthy-speech codec benchmarks. When trained separately on each healthcare dataset, performance improves but remains limited, suggesting that more informative representations and mechanisms for capturing heterogeneous cues are still needed. To broaden this comparison, we additionally evaluate RawNet2 \cite{9414234}, LCNN \cite{wu20c_interspeech}, and SAMO \cite{10094704}, representing waveform-based, spectrogram-based, and robust generalization-oriented detector families. Although these baselines provide modest improvements over the weakest transfer setting, they still remain clearly below the stronger PTM-based approaches reported next. The AASIST variant equipped with a wav2vec~2.0 backbone is also consistently stronger than the standard AASIST settings, but still leaves a substantial gap to the best-performing methods introduced later. Taken together, these results show that the challenge is not specific to a single detector architecture, but reflects a broader difficulty of transferring existing codec-fake detectors to pathological speech.

\noindent \textbf{PTM comparison with FCN and CNN as downstream networks}: We next benchmark a diverse set of pre-trained encoders to assess which representations are most effective for healthcare codec-deepfake detection. Using the same train/test protocol, we train two lightweight downstream heads (FCN and 1D-CNN) on top of each frozen PTM embedding, and report results in Table~\ref{tab:ptm_clinical_results} for both English and Chinese subsets. Two consistent observations emerge. First, the CNN head—despite its simplicity—yields stronger performance than the FCN in most settings, indicating that local temporal structure remains informative and can be exploited with a shallow convolutional frontend. Second, among all upstream encoders, PaSST provides the strongest single-representation baseline across tasks and languages, achieving the best overall performance in both English and Chinese. We further observe that Alzheimer’s is consistently more challenging than the other clinical tasks for all PTMs, suggesting that condition-induced variability and broader acoustic differences make codec artifacts harder to isolate in this setting. Finally, performance is generally lower on the Chinese subset than on English for the same configuration, highlighting an additional cross-lingual shift that compounds the healthcare-domain mismatch. Notably, the comparatively strong results of Whisper across tasks align with prior findings that multilingual speech foundation models often provide more robust features for deepfake-related tasks than monolingual SSL encoders, likely due to broader linguistic diversity during pre-training \cite{phukan2025svdsa}. \newline
\begin{table}[!hbt]
\centering
\scriptsize
\setlength{\tabcolsep}{6pt}
\renewcommand{\arraystretch}{1.15}
\begin{tabular}{l|cc|cc|cc}
\toprule
\multicolumn{1}{c|}{\multirow{2}{*}{\textbf{PTMs}}} &
\multicolumn{2}{c|}{\textbf{Dep}} &
\multicolumn{2}{c|}{\textbf{Alz}} &
\multicolumn{2}{c}{\textbf{Dys}} \\
\cmidrule(lr){2-3}\cmidrule(lr){4-5}\cmidrule(lr){6-7}
\multicolumn{1}{c|}{} &
\textbf{Acc $\uparrow$} & \textbf{F1 $\uparrow$} &
\textbf{Acc $\uparrow$} & \textbf{F1 $\uparrow$} &
\textbf{Acc $\uparrow$} & \textbf{F1 $\uparrow$} \\
\midrule
\multicolumn{7}{c}{\textbf{ENGLISH}} \\
\midrule
WLM    & 94.27 & 91.52 & 92.84 & 91.56 & 95.07 & 92.67 \\
WV2 & 95.68 & 93.14 & 95.39 & 92.78 & 94.82 & 93.19 \\
WHI  & 94.12 & 92.38 & 93.11 & 91.34 & 95.43 & 92.82 \\
XVE  & 94.46 & 91.59 & 91.65 & 90.07 & 93.24 & 91.54 \\
PST    & 97.04 & 96.81 & 96.73 & 95.20 & 96.57 & 94.28 \\
\midrule
\multicolumn{7}{c}{\textbf{CHINESE}} \\
\midrule
WLM    & 91.56 & 89.74 & 89.25 & 86.62 & 91.48 & 90.29 \\
WV2 & 93.04 & 90.42 & 93.08 & 91.54 & 92.03 & 89.47 \\
WHI  & 90.29 & 88.19 & 90.71 & 89.03 & 93.87 & 91.86 \\
XVE  & 92.17 & 89.87 & 87.93 & 84.72 & 89.61 & 86.95 \\
PST    & 94.41 & 92.10 & 94.40 & 92.18 & 93.20 & 91.42 \\
\bottomrule
\end{tabular}
\caption{\textbf{PHOENIX-Mamba} results. Abbreviations follow Table~\ref{tab:ptm_clinical_results}.}
\label{mambaphoenix}
\vspace{-0.3cm}
\end{table}

\begin{table*}[!hbt]
\centering
\scriptsize
\setlength{\tabcolsep}{6pt}
\renewcommand{\arraystretch}{1.1}
\begin{tabular}{l|cccccc|cccccc}
\toprule
& \multicolumn{6}{c|}{\textbf{English}} & \multicolumn{6}{c}{\textbf{Chinese}} \\
\cmidrule(lr){2-7} \cmidrule(lr){8-13}
\textbf{PTMs} 
& \multicolumn{2}{c}{\textbf{Dep}} 
& \multicolumn{2}{c}{\textbf{Alz}} 
& \multicolumn{2}{c|}{\textbf{Dys}} 
& \multicolumn{2}{c}{\textbf{Dep}} 
& \multicolumn{2}{c}{\textbf{Alz}} 
& \multicolumn{2}{c}{\textbf{Dys}} \\
\cmidrule(lr){2-3} \cmidrule(lr){4-5} \cmidrule(lr){6-7}
\cmidrule(lr){8-9} \cmidrule(lr){10-11} \cmidrule(lr){12-13}
& \textbf{Acc$\uparrow$} & \textbf{F1$\uparrow$} 
& \textbf{Acc$\uparrow$} & \textbf{F1$\uparrow$} 
& \textbf{Acc$\uparrow$} & \textbf{F1$\uparrow$} 
& \textbf{Acc$\uparrow$} & \textbf{F1$\uparrow$} 
& \textbf{Acc$\uparrow$} & \textbf{F1$\uparrow$} 
& \textbf{Acc$\uparrow$} & \textbf{F1$\uparrow$} \\
\midrule
WavLM    
& 91.43 & 89.14 & 87.65 & 85.98 & 91.32 & 89.42 
& 89.62 & 88.74 & 85.45 & 83.73 & 88.21 & 86.92 \\

Wav2Vec2 
& 91.67 & 88.34 & 92.82 & 90.38 & 91.49 & 87.04 
& 88.45 & 87.64 & 90.92 & 88.69 & 89.82 & 87.54 \\

Whisper  
& 89.51 & 87.93 & 91.42 & 88.61 & 92.74 & 89.74 
& 90.04 & 87.61 & 89.54 & 87.94 & 90.05 & 88.59 \\

X-vector 
& 90.37 & 88.52 & 85.27 & 83.74 & 88.49 & 87.29 
& 89.75 & 86.05 & 87.72 & 85.82 & 90.92 & 89.34 \\

PaSST    
& \textbf{95.59} & \textbf{93.29} 
& \textbf{94.66} & \textbf{92.04} 
& \textbf{95.17} & \textbf{93.28} 
& \textbf{94.17} & \textbf{93.74} 
& \textbf{95.09} & \textbf{92.75} 
& \textbf{93.99} & \textbf{93.02} \\
\bottomrule
\end{tabular}
\caption{Seen--unseen codec evaluation results.}
\label{tab:seenunseencodec}
\end{table*}

\noindent \textbf{Results of \textbf{\texttt{PHOENIX-Mamba}}}: To understand where the improvements come from, we compare \textbf{\texttt{PHOENIX-Mamba}} against the strongest single-representation baselines reported in Table~\ref{tab:ptm_clinical_results} under the same train/test protocol. The results in Table~\ref{mambaphoenix} show that replacing standard single-vector classification with our evidence-driven, prototype-based reasoning consistently improves performance across all upstream encoders, tasks, and languages, indicating that the gains are primarily driven by the proposed modeling strategy rather than any single PTM. With PaSST, \textbf{\texttt{PHOENIX-Mamba}} achieves the best overall performance, reaching 97.04/96.81 on Dep, 96.73/95.20 on Alz, and 96.57/94.28 on Dys for the English subset (Acc/F1), and maintaining strong results on the Chinese subset with 94.41/92.10 (Dep), 94.40/92.18 (Alz), and 93.20/91.42 (Dys). Notably, the largest gains appear on Alzheimer’s, where variability is highest, reinforcing the importance of explicitly handling heterogeneity in healthcare speech. The consistent gains suggest that \textbf{\texttt{PHOENIX-Mamba}} is effective because it aligns the decision process with the structure of the problem—localized cues and heterogeneous artifact modes—rather than relying on a single pooled representation. 

To complement Accuracy and macro-F1, we also report EER, a standard threshold-independent metric in synthetic speech detection \cite{sheth-etal-2025-curved}. Using the same PaSST upstream representation, the CNN baseline attains EERs of 14.01/16.52/15.93 on English Dep/Alz/Dys and 18.42/20.04/21.78 on Chinese, whereas \textbf{\texttt{PHOENIX-Mamba}} achieves markedly lower values of 5.17/6.29/6.23 and 6.54/5.42/6.79, respectively. This further confirms the advantage of the proposed method under the same evaluation setting.

\noindent \textbf{{Generalization Beyond the In-Domain Setting}:} We also study the generalization behavior of \textbf{\texttt{PHOENIX-Mamba}} under a held-out codec-family protocol. For this evaluation, the seven codec families in HCFK are partitioned into seen and unseen subsets by randomly assigning five families to training and holding out the remaining two exclusively for testing. As shown in Table \ref{tab:seenunseencodec}, the relative behavior across upstream encoders remains largely stable in this more challenging setting, with PaSST continuing to achieve the strongest overall performance. This suggests that the proposed framework captures evidence that remains informative even for codec families not encountered during training.

To further examine transfer beyond codec variation, we also evaluated \textbf{\texttt{PHOENIX-Mamba}} under a cross-pathology setting in which training and testing were performed on different clinical conditions. When trained on Depression and evaluated on Alzheimer’s, \textbf{\texttt{PHOENIX-Mamba}} achieved 95.88 Acc / 93.41 F1 / 6.57 EER in English and 91.79 Acc / 89.05 F1 / 6.82 EER in Chinese. Under a broader transfer setting, where training was performed on Depression and Dysarthria and evaluation was conducted on Alzheimer’s, performance further improved to 98.53 Acc / 97.21 F1 / 3.66 EER in English and 97.84 Acc / 95.10 F1 / 3.79 EER in Chinese. These results indicate that \textbf{\texttt{PHOENIX-Mamba}} retains transferable codec-fake evidence even when the target condition differs from those observed during training, and that broader pathology coverage during training further strengthens generalization.
\begin{table}[!hbt]
\centering
\scriptsize
\setlength{\tabcolsep}{2.5pt}
\renewcommand{\arraystretch}{1.05}
\begin{tabular}{l|cc|cc|cc}
\toprule
\multicolumn{1}{c|}{\multirow{2}{*}{\textbf{Methods}}} &
\multicolumn{2}{c|}{\textbf{Dep}} &
\multicolumn{2}{c|}{\textbf{Alz}} &
\multicolumn{2}{c}{\textbf{Dys}} \\
\cmidrule(lr){2-3}\cmidrule(lr){4-5}\cmidrule(lr){6-7}
\multicolumn{1}{c|}{} &
\textbf{Acc} & \textbf{F1} &
\textbf{Acc} & \textbf{F1} &
\textbf{Acc} & \textbf{F1} \\
\midrule
\multicolumn{7}{c}{\textbf{ENGLISH}} \\
\midrule
CNN Head                           & 82.26 & 80.73 & 75.52 & 72.13 & 79.37 & 77.91 \\
BiGRU Head                         & 87.69 & 84.91 & 82.86 & 80.49 & 86.61 & 83.73 \\
Single evidence: set $M{=}1$       & 73.51 & 72.02 & 55.03 & 52.67 & 67.94 & 65.02 \\
PHOENIX-Euc & 83.62 & 81.24 & 79.48 & 77.16 & 84.72 & 83.67 \\
PHOENIX-Mamba (Full)           & 97.04 & 94.81 & 96.73 & 94.20 & 96.05 & 93.28 \\
\midrule
\multicolumn{7}{c}{\textbf{CHINESE}} \\
\midrule
CNN Head                           & 79.81 & 77.26 & 73.12 & 70.46 & 76.89 & 74.28 \\
BiGRU Head                         & 83.62 & 80.91 & 78.95 & 77.87 & 84.03 & 81.65 \\
Single evidence: set $M{=}1$       & 71.04 & 69.38 & 52.46 & 50.24 & 65.26 & 63.77 \\
PHOENIX-Euc & 80.59 & 77.14 & 77.08 & 74.63 & 80.91 & 77.98 \\
PHOENIX-Mamba (Full)           & 94.41 & 92.10 & 94.40 & 92.18 & 93.20 & 91.42 \\
\bottomrule
\end{tabular}
\caption{Ablation study of \textbf{\texttt{PHOENIX-Mamba}}}
\label{tab:ablation_phoenix}
\vspace{-0.3cm}
\end{table}

\subsection{Ablation Study}
To quantify the contribution of key design choices in \textbf{\texttt{PHOENIX-Mamba}}, we perform an ablation study along three axes, keeping the training protocol fixed and varying a single component at a time (Table~\ref{tab:ablation_phoenix}).

\noindent \textbf{Role of Temporal Modeling:} We first examine the impact of the sequence modeling head used for downstream reasoning. Starting from a lightweight CNN head, we replace it with a stronger recurrent alternative (BiGRU head) while keeping the remaining pipeline unchanged. This comparison isolates the benefit of richer temporal dependencies over shallow local modeling. \newline
\noindent \textbf{Role of Multi-Evidence Pooling}
A central design choice in \textbf{\texttt{PHOENIX-Mamba}} is to retain multiple localized cues via $M$ evidence vectors rather than collapsing an utterance into a single summary. To isolate this effect, we reduce the evidence set to a single vector by setting $M{=}1$ (single-evidence variant), while keeping the backbone and classifier identical. This ablation tests whether preserving multiple evidences is necessary for healthcare speech, where codec artifacts may appear intermittently and non-uniformly. \newline
\noindent \textbf{Role of Geometry-Aware Multi-Mode Reasoning}
Finally, we evaluate the importance of the geometry-aware prototype reasoning used to model heterogeneous fake modes. We compare the full \textbf{\texttt{PHOENIX-Mamba}} to a Euclidean counterpart (PHOENIX-Euc) that removes the hyperbolic embedding/clustering component while retaining the same overall architecture and optimization setup. This variant isolates the contribution of the manifold-based multi-mode structure beyond standard Euclidean classification.

\section{Conclusion}
In this work, we initiate a focused study of Healthcare CodecFake Detection (HCFD), targeting codec-generated audio deepfakes in pathological speech—a setting that remains underexplored despite its practical relevance to clinical and telehealth communication. To support systematic evaluation, we introduce Healthcare CodecFake (HCFK), a benchmark constructed by resynthesizing pathological speech across multiple clinical conditions and two languages using a diverse set of neural audio codecs. Our findings show that detectors trained under standard codec-deepfake benchmarks exhibit limited transfer to healthcare speech, highlighting a substantial domain shift and motivating dedicated solutions for this problem. Building on this insight, we benchmark a range of pre-trained audio/speech encoders and observe that representation quality and downstream modeling both play an important role. To move beyond single-vector classification, we propose \textbf{\texttt{PHOENIX-Mamba}}, a geometry-aware framework that retains multiple localized evidences and models the fake class through multiple prototype modes in hyperbolic space. Across clinical tasks and languages, \textbf{\texttt{PHOENIX-Mamba}} consistently improves over strong PTM baselines, and ablations confirm the importance of multi-evidence pooling and geometry-aware multi-mode reasoning. We expect HCFK and \textbf{\texttt{PHOENIX-Mamba}} to provide a foundation for more reliable evaluation and continued progress on codec-deepfake detection in healthcare-oriented speech.

\section*{Limitations \& Future Work}
\textbf{Limitations} HCFK provides a first pathology-aware benchmark for healthcare codec-fake detection, but currently covers a limited set of conditions and languages. We focus on codec-based resynthesis with a fixed set of NACs under a controlled protocol; broader real-world channel effects and other attack families (e.g., TTS/VC/diffusion, replay/recapture, adversarial post-processing) are not included. We evaluate detection only and do not address generator/codec attribution or open-set uncertainty for unseen attacks. \newline
\noindent \textbf{Future work} We will expand coverage to more conditions, languages, and recording scenarios, including telehealth-style settings, and extend the benchmark to additional attack families beyond NAC resynthesis. We also plan to study open-set detection, uncertainty estimation, and attribution, as well as privacy-preserving evaluation and more interpretable artifact analysis.

\section*{Ethical considerations}
This work is motivated by the need to safeguard clinical and healthcare communication against codec-generated audio manipulation. We do not collect any new human-subject recordings. The benchmark is constructed by applying NACs to existing speech datasets that are available for research use under their respective licenses and access conditions. As an additional quality-control measure, a certified speech therapist with clinical experience in pathological speech assessment qualitatively reviewed a subset of bona fide and codec-synthesized samples. The proposed models and benchmarks are not intended for medical diagnosis, treatment decisions, or deployment as standalone security mechanisms in clinical workflows.

\section*{Acknowledgments}
The authors gratefully acknowledge the support of the United States--Ireland--Northern Ireland R\&D Partnership Programme (USI-207), and access to the Tier 2 High-Performance Computing resources provided by the Northern Ireland High Performance Computing (NIHPC) facility, funded by the UK Engineering and Physical Sciences Research Council (EPSRC), Grant No. EP/T022175/1.

\bibliography{custom}

\appendix
\section*{Appendix}

In the appendix, we provide:
\begin{itemize}
    \item \textbf{Section A: Neural Audio Codecs.} Details of the neural audio codec families and the specific publicly released checkpoints used in our experiments.
    \item \textbf{Section B: Hyperparameters and System Configurations.} A summary of key training hyperparameters and geometry-related settings (Table~\ref{tab:hyperparams_hyperbolic_clusters}).
    \item \textbf{Section C: Visualization Analysis.} Additional qualitative analyses including confusion matrices (Figure~\ref{fig:cm_8}) and t-SNE plots (Figure~\ref{fig:tsne_8}) for representative configurations.
\end{itemize}

\section{Neural Audio Codecs}
\label{sec:neural_audio_codecs}

Following \citet{lu24f_interspeech} and \citet{wu24p_interspeech}, we use the same family of neural audio codecs, focusing on state-of-the-art, publicly released models that are easy to reproduce and widely used.

\noindent \textbf{Speechtokenizer \cite{zhang2024speechtokenizer}\footnote{\url{https://github.com/ZhangXInFD/SpeechTokenizer.git}}:}
It is a unified speech tokenizer built on an RVQ-GAN style neural codec. The model uses an EnCodec-based convolutional encoder--decoder backbone with Residual Vector Quantization (RVQ). We use the 16~kHz SpeechTokenizer setting in this work.

\noindent \textbf{Descript Audio Codec \cite{kumar2024high}\footnote{\url{https://huggingface.co/descript/dac_16khz}}:}
It is a VQ-GAN-based neural audio codec targeting high-fidelity reconstruction. The approach discretizes encoder features with RVQ and trains the generator using adversarial learning alongside multi-scale frequency-domain criteria to suppress codec artifacts. We evaluate its checkpoints at 16~kHz, 24~kHz, and 44~kHz sampling rates.

\noindent \textbf{Encodec \cite{defossez2022high}\footnote{\url{https://huggingface.co/facebook/encodec_24khz}}:}
It is a streaming, high-quality neural codec that couples a convolutional encoder--decoder with RVQ discretization. Training combines time-domain and frequency-domain reconstruction criteria with a spectrogram adversary for improved perceptual quality. We evaluate the 24~kHz and 48~kHz models in our study.

\noindent \textbf{Soundstream \cite{zeghidour2021soundstream}\footnote{\url{https://github.com/haydenshively/SoundStream}}:}
It is an end-to-end neural codec tailored for low-bitrate speech compression. The model combines an encoder--decoder backbone with Residual Vector Quantization (RVQ) and multi-scale STFT discriminators, enabling high perceptual quality under aggressive compression (3--18~kbps). We adopt the 16~kHz variant in our experiments (\texttt{soundstream\_16khz}).

\noindent \textbf{Funcodec \cite{du2024funcodec}\footnote{\url{https://github.com/modelscope/FunCodec}}:}
It is an open-source neural speech codec toolkit built to make modern codec models easy to train, reproduce, and integrate into downstream pipelines. It extends the FunASR ecosystem and provides unified training recipes and inference scripts for widely used neural codec families.

\noindent \textbf{Audiodec \cite{wu2023audiodec}\footnote{\url{https://github.com/facebookresearch/AudioDec}}:}
It is a high-quality neural codec formulated as an end-to-end autoencoder. Training proceeds in two phases: it first learns the encoder--decoder with metric-based objectives to ensure stable optimization, and then applies an adversarial refinement stage that updates only the decoder to enhance waveform realism. In our experiments, we use the 28~kHz and 48~kHz AudioDec variants.

\noindent \textbf{SNAC \cite{siuzdak2024snac}\footnote{\url{https://huggingface.co/hubertsiuzdak/snac_44khz}}:}
It is a multi-scale neural audio codec that extends standard Residual Vector Quantization by allowing different quantizers to operate at different temporal resolutions. Concretely, it forms a hierarchical token representation by quantizing coarse-to-fine structure at multiple frame rates. In this study, we employ various sampling rates, specifically 24~kHz, 32~kHz, and 44~kHz.

\section{Hyperparameters and System Configurations}
\label{sec:hyperparameters}

Table~\ref{tab:hyperparams_hyperbolic_clusters} summarizes the key hyperparameters used in all experiments, including the Poincar\'e-ball geometry settings, VQ and HEL configurations, OT/Sinkhorn parameters, and optimization details.

\begin{table}[!hbt]
\centering
\setlength{\tabcolsep}{4pt}
\renewcommand{\arraystretch}{1.15}
\begin{tabular}{@{}l| l@{}}
\hline
\textbf{Hyperparameter} & \textbf{Value} \\
\hline
\multicolumn{2}{@{}l}{\textit{Geometry / Manifold (Poincar\'e ball)}}\\
Hyperbolic curvature & $\kappa=-1.0$ \\
Hyperbolic embedding dim & $h = 128$ \\
\hline
\multicolumn{2}{@{}l}{\textit{Sequence \& Evidence Pooling}}\\
Adapter output dim & $d = 256$ \\
\# evidence vectors (glimpses) & $M = 4$ \\
\hline
\multicolumn{2}{@{}l}{\textit{Binary Prototypes (self-discovered positive modes)}}\\
Negative prototypes & $|p_-|=1$ \\
\# positive prototypes (modes) & $K = 4$ \\
Temperature (assignments / soft-min) & $\tau = 0.1$ \\
\hline
\multicolumn{2}{@{}l}{\textit{Loss Weights}}\\
Cluster loss weight & $\lambda = 1.0$ \\
Separation (repulsion) weight & $\beta = 0.1$ \\
Entropy regularizer weight & $\gamma = 0.05$ \\
\hline
\multicolumn{2}{@{}l}{\textit{Optimization}}\\
Optimizer & AdamW \\
AdamW betas & $(0.9, 0.999)$ \\
AdamW epsilon & $10^{-8}$ \\
Weight decay & $0.01$ \\
LR (encoder, if finetuned) & $3\times 10^{-5}$ \\
LR (new layers: $\phi, f_\theta, W$, prototypes) & $1\times 10^{-4}$ \\
Gradient clipping & $1.0$ \\
Epochs & $20$ \\
Batch size & $|\mathcal{B}|=32$ \\
\hline
\end{tabular}
\caption{Hyperparameters for geometry-aware sequence classification with self-discovered hyperbolic clusters. Shared default configuration used across all main experiments.}
\label{tab:hyperparams_hyperbolic_clusters}
\end{table}

\section{Visualization Analysis}
\label{sec:visualization_analysis}

\subsection{Confusion Matrices}
\begin{figure*}[!hbt]
    \centering
    \subfloat[]{\includegraphics[width=0.32\textwidth]{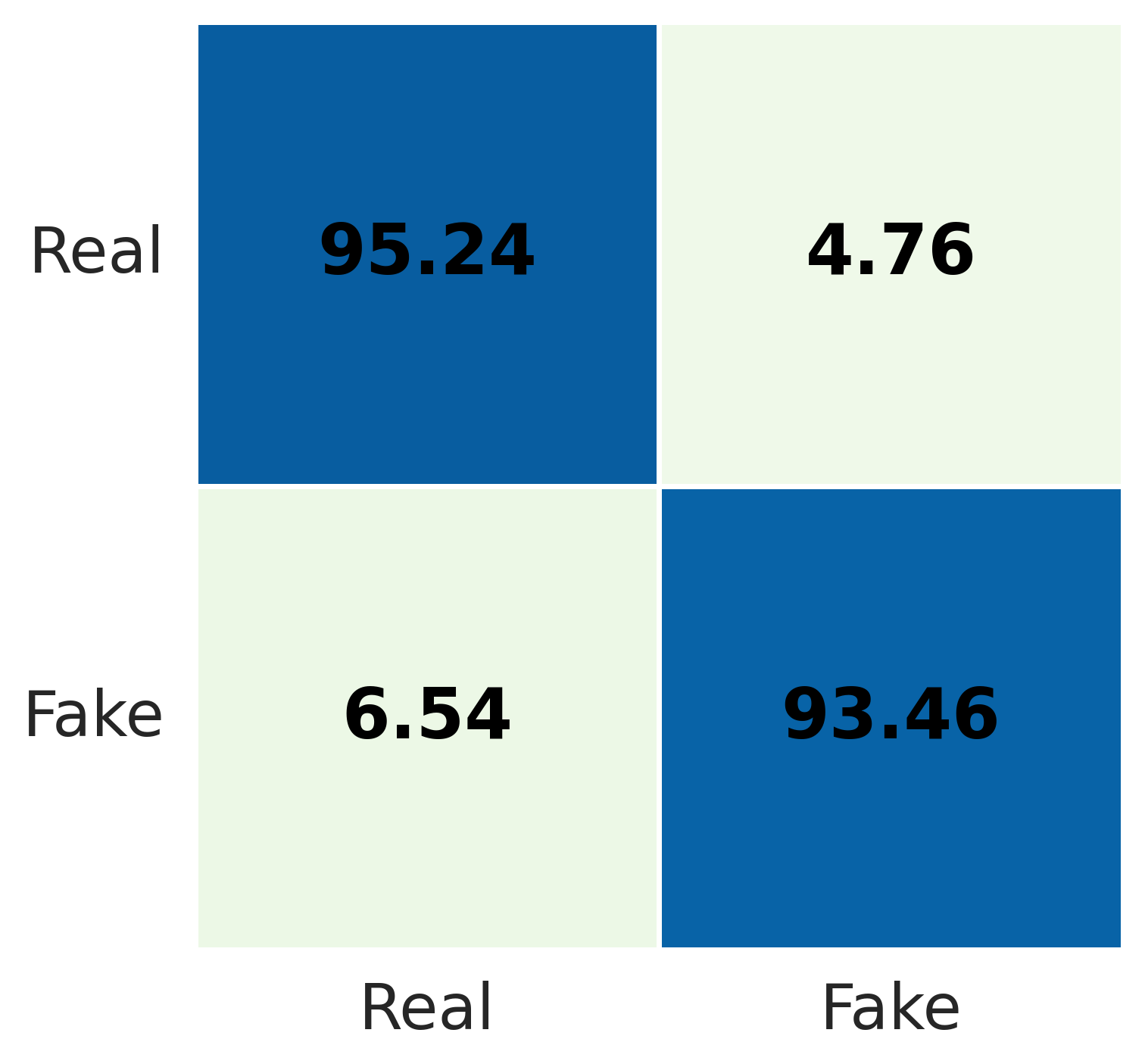}}
    \hspace{0.03\textwidth}
    \subfloat[]{\includegraphics[width=0.32\textwidth]{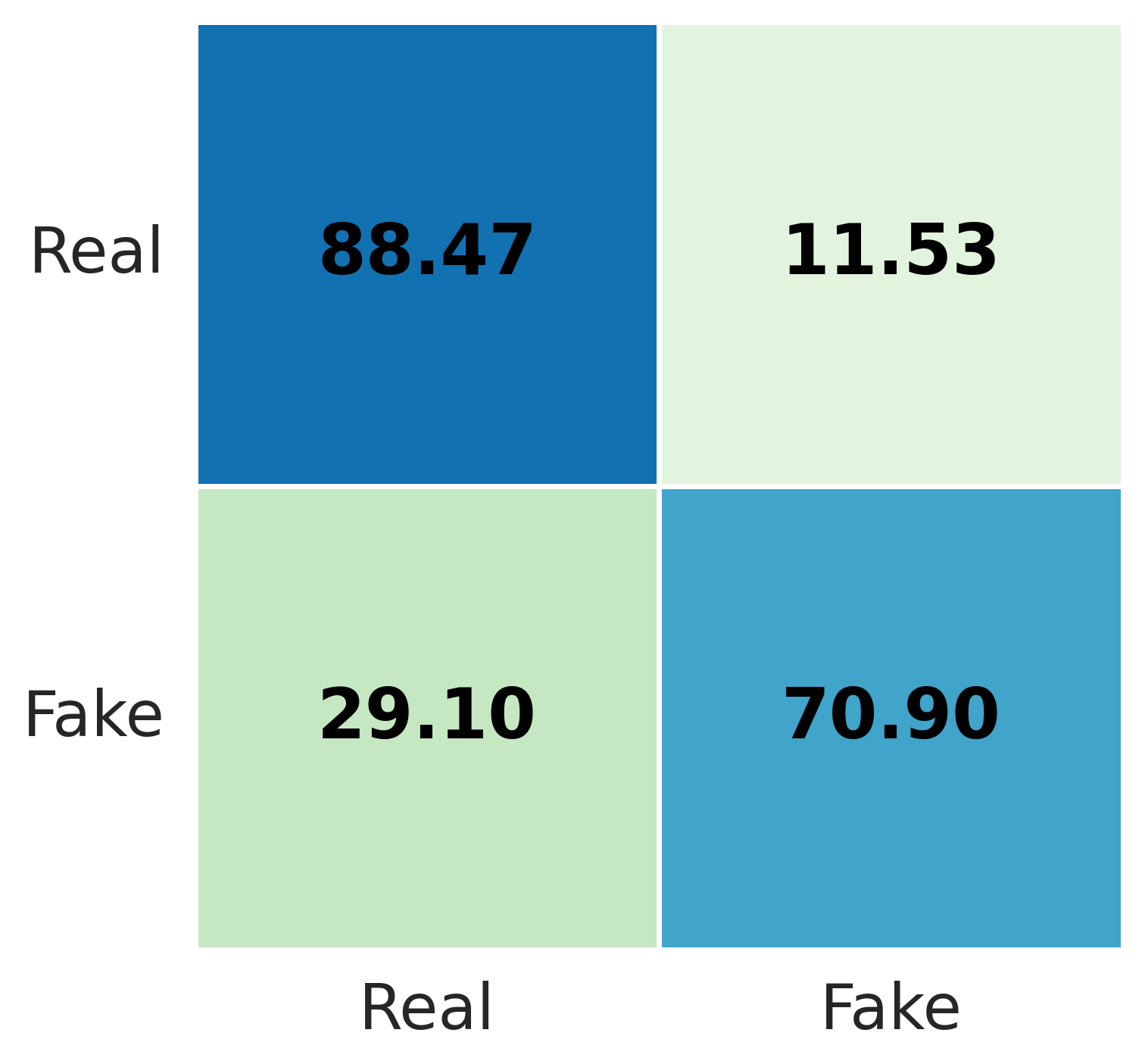}}\\[1ex]

    \subfloat[]{\includegraphics[width=0.32\textwidth]{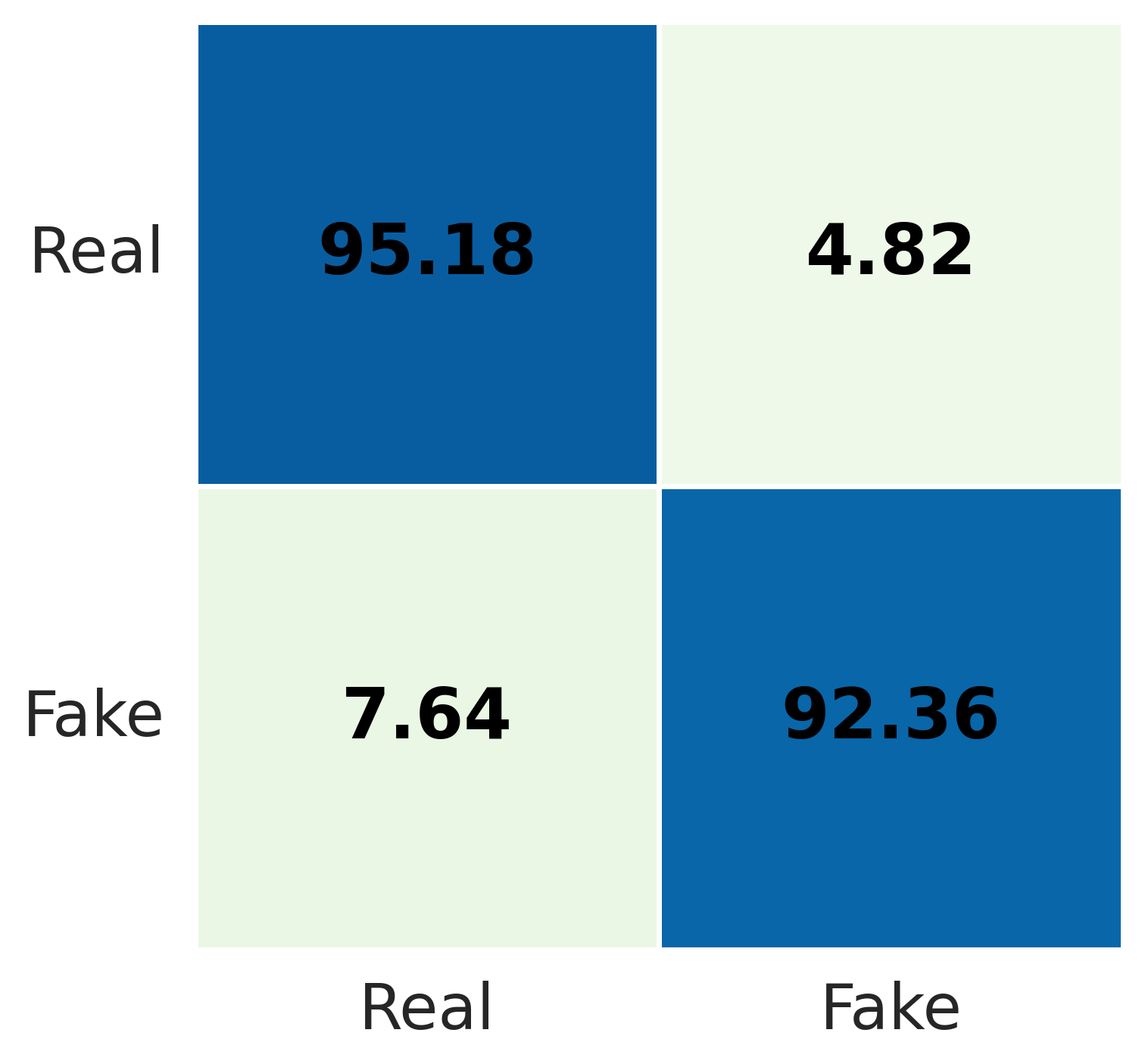}}
    \hspace{0.03\textwidth}
    \subfloat[]{\includegraphics[width=0.32\textwidth]{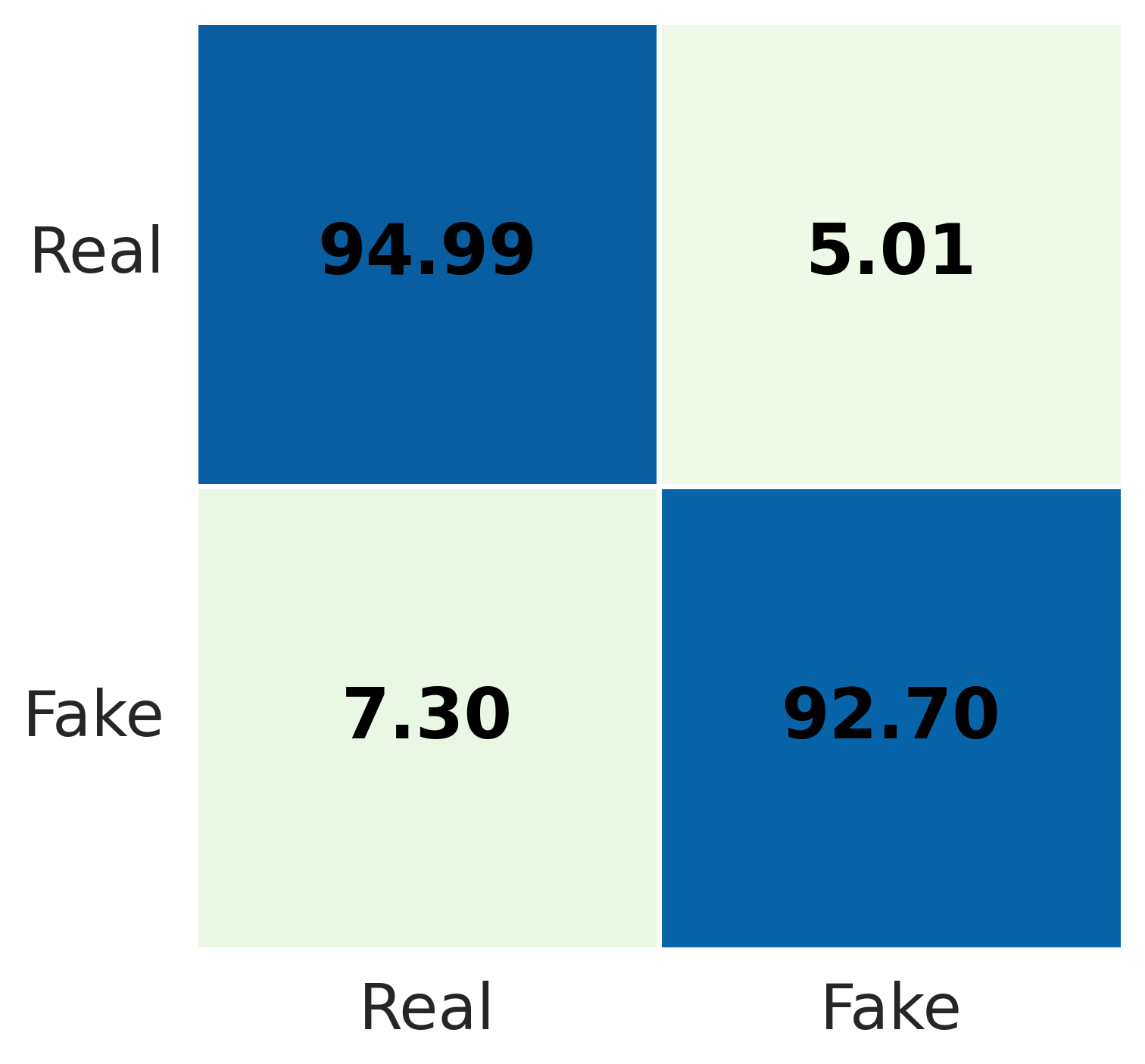}}\\[1ex]

    \subfloat[]{\includegraphics[width=0.32\textwidth]{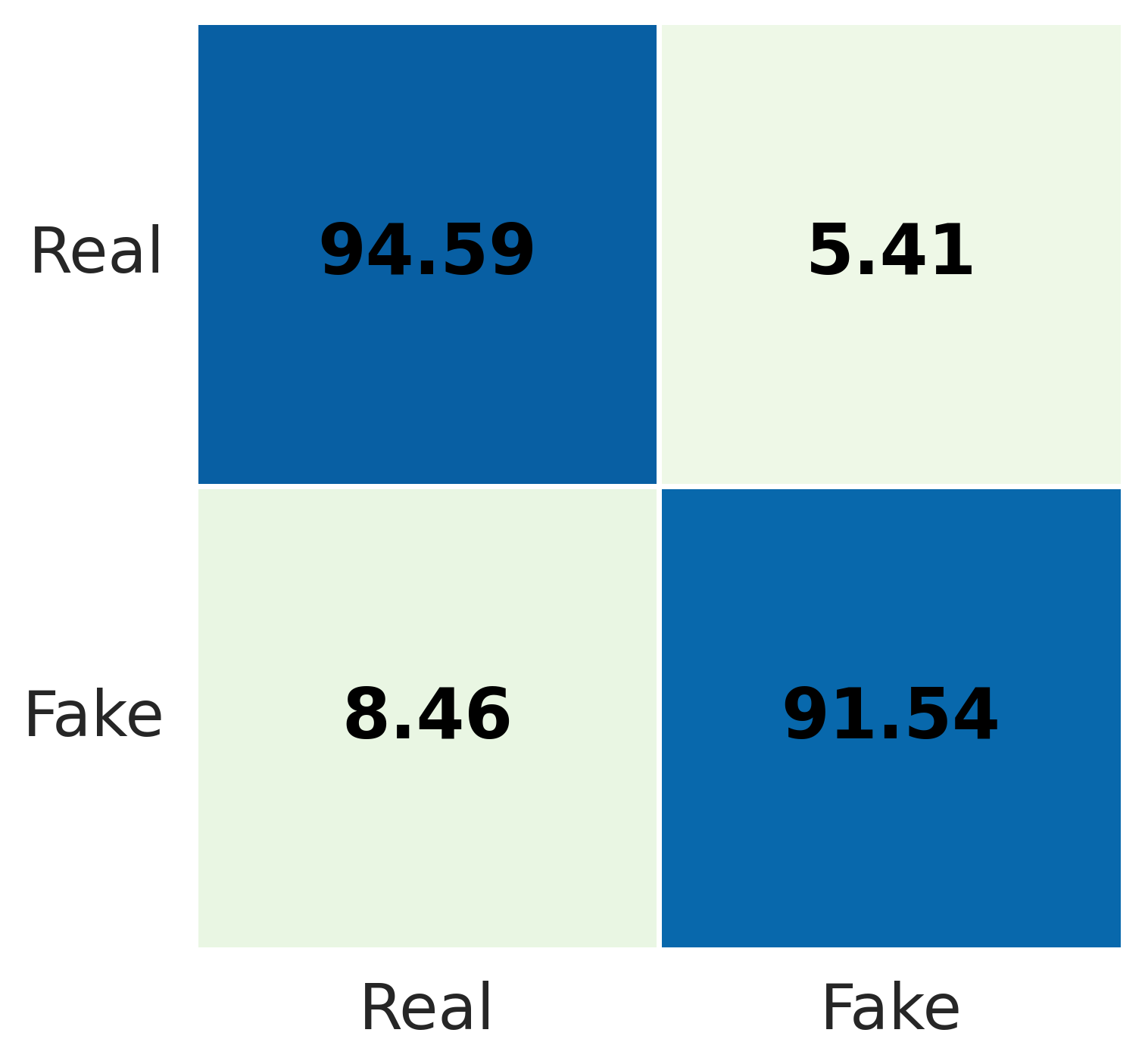}}
    \hspace{0.03\textwidth}
    \subfloat[]{\includegraphics[width=0.32\textwidth]{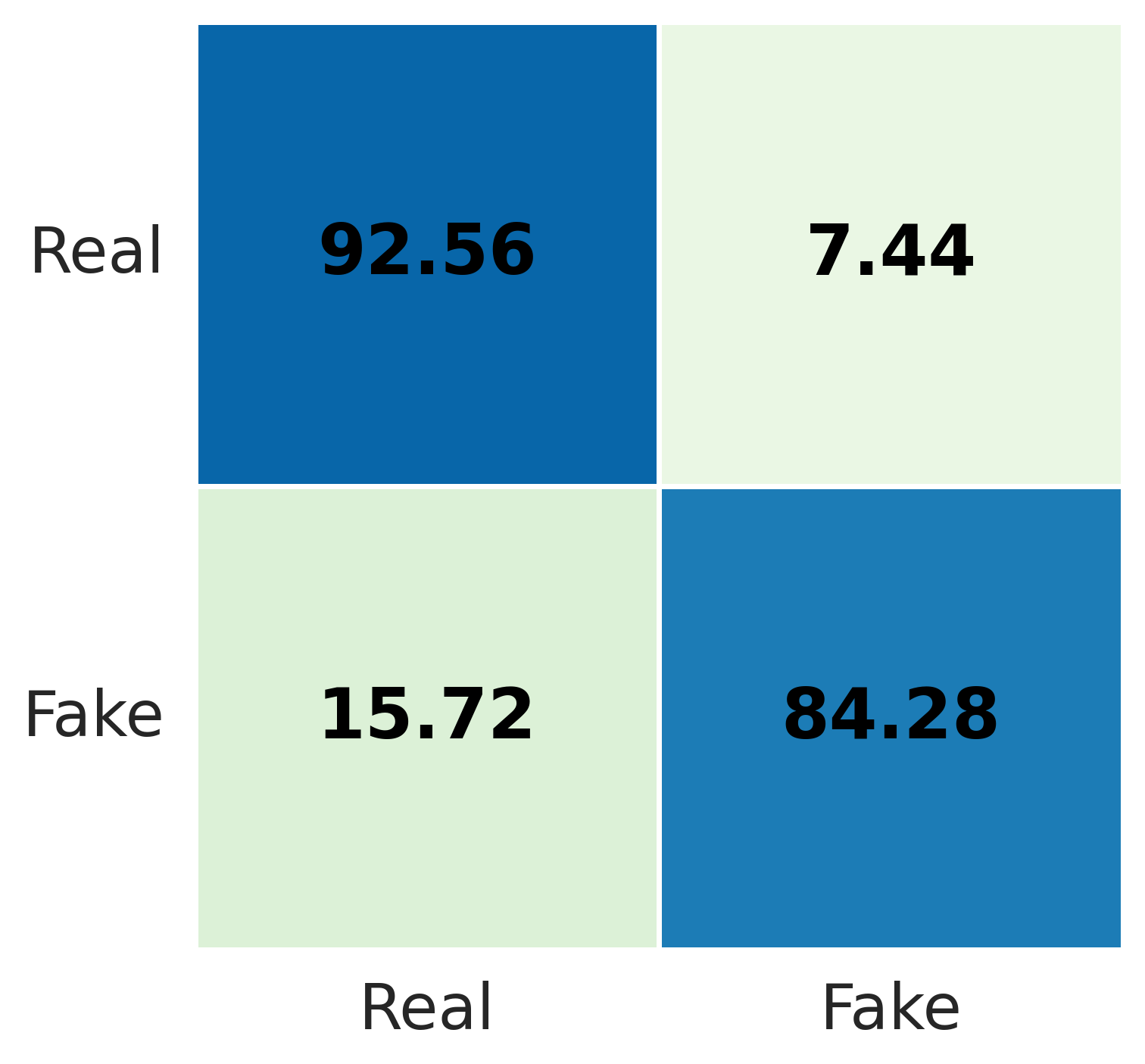}}\\[1ex]

    \caption{Confusion matrices for selected \texttt{PHOENIX-Mamba} configurations: (a) Depression with PaSST on Chinese; (b) Depression with Wav2vec~2.0 on English; (c) Dysarthria with PaSST on Chinese; (d) Alzheimer’s with PaSST on Chinese; (e) Alzheimer’s with Whisper on Chinese; and (f) Alzheimer’s with WavLM on Chinese. These plots summarize prediction stability and highlight the dominant error modes for each configuration.}
    \label{fig:cm_8}
\end{figure*}

Figure~\ref{fig:cm_8} provides confusion matrices for representative \texttt{PHOENIX-Mamba} configurations across tasks, PTM backbones, and languages. These plots offer a class-wise view of where the model is most reliable and where errors concentrate, complementing the aggregate Acc/F1 trends reported in the main results.

\subsection{t-SNE Plots}
\begin{figure*}[!hbt]
    \centering
    \subfloat[]{\includegraphics[width=0.42\textwidth]{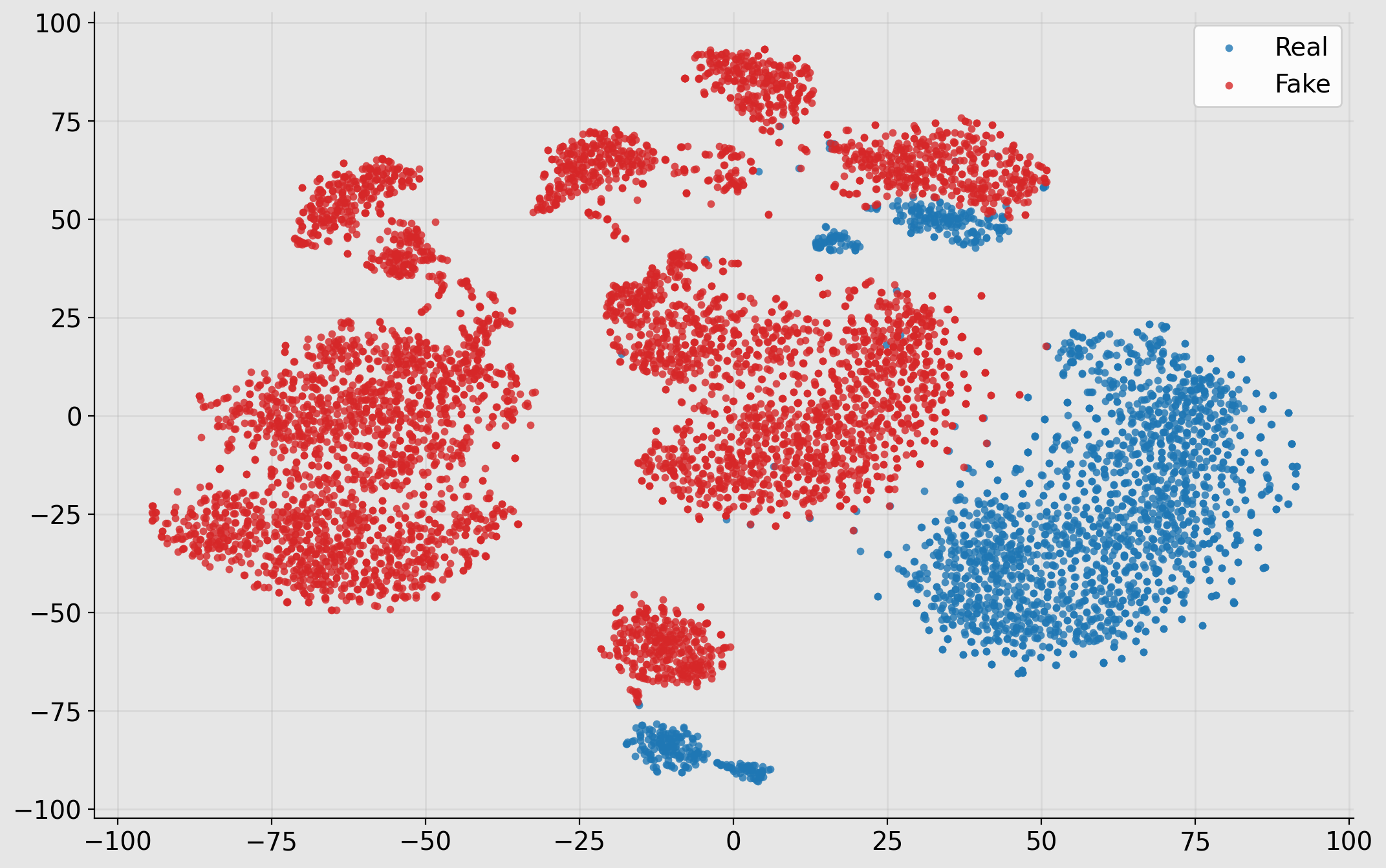}}
    \hspace{0.03\textwidth}
    \subfloat[]{\includegraphics[width=0.42\textwidth]{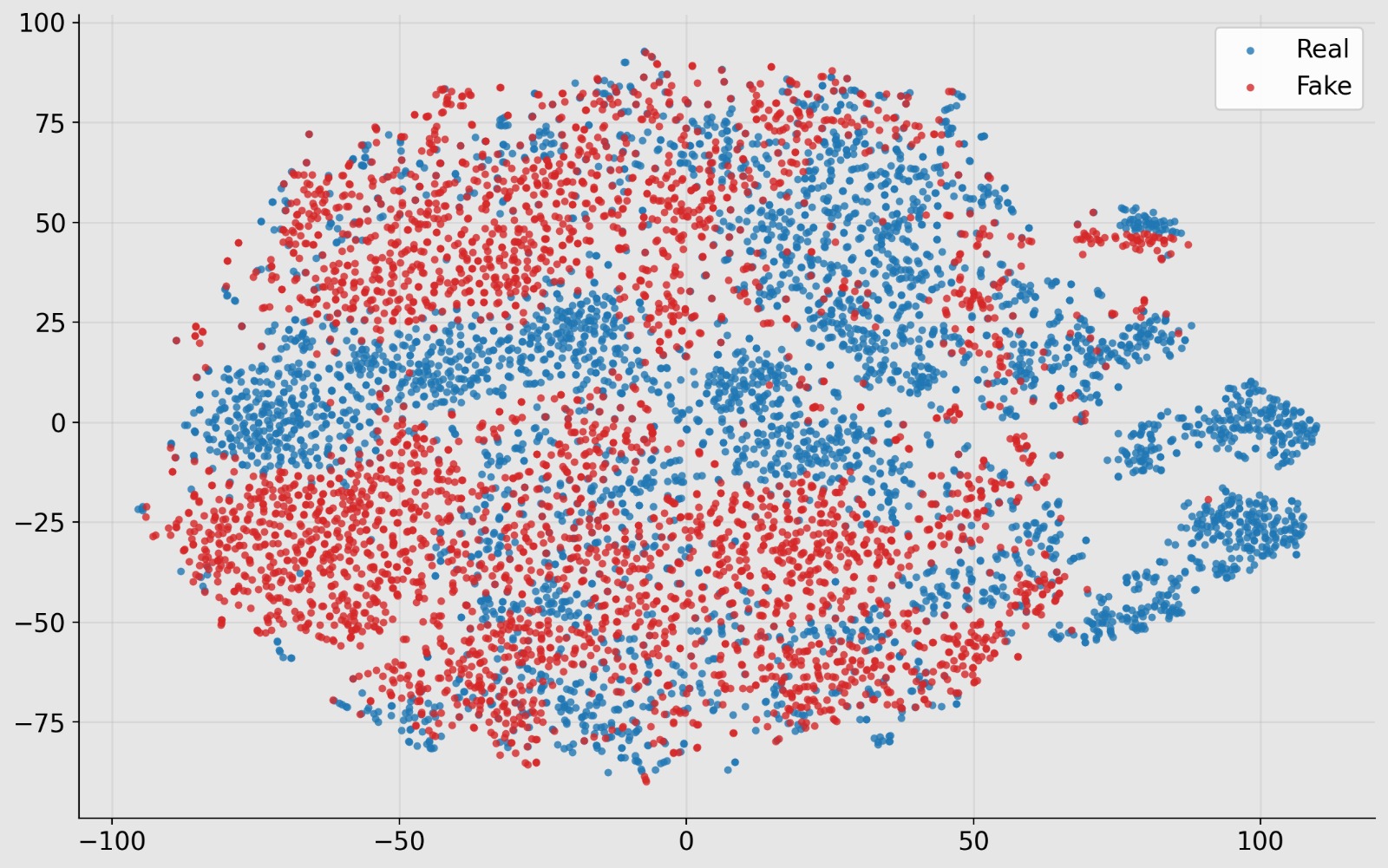}}\\[1ex]

    \subfloat[]{\includegraphics[width=0.42\textwidth]{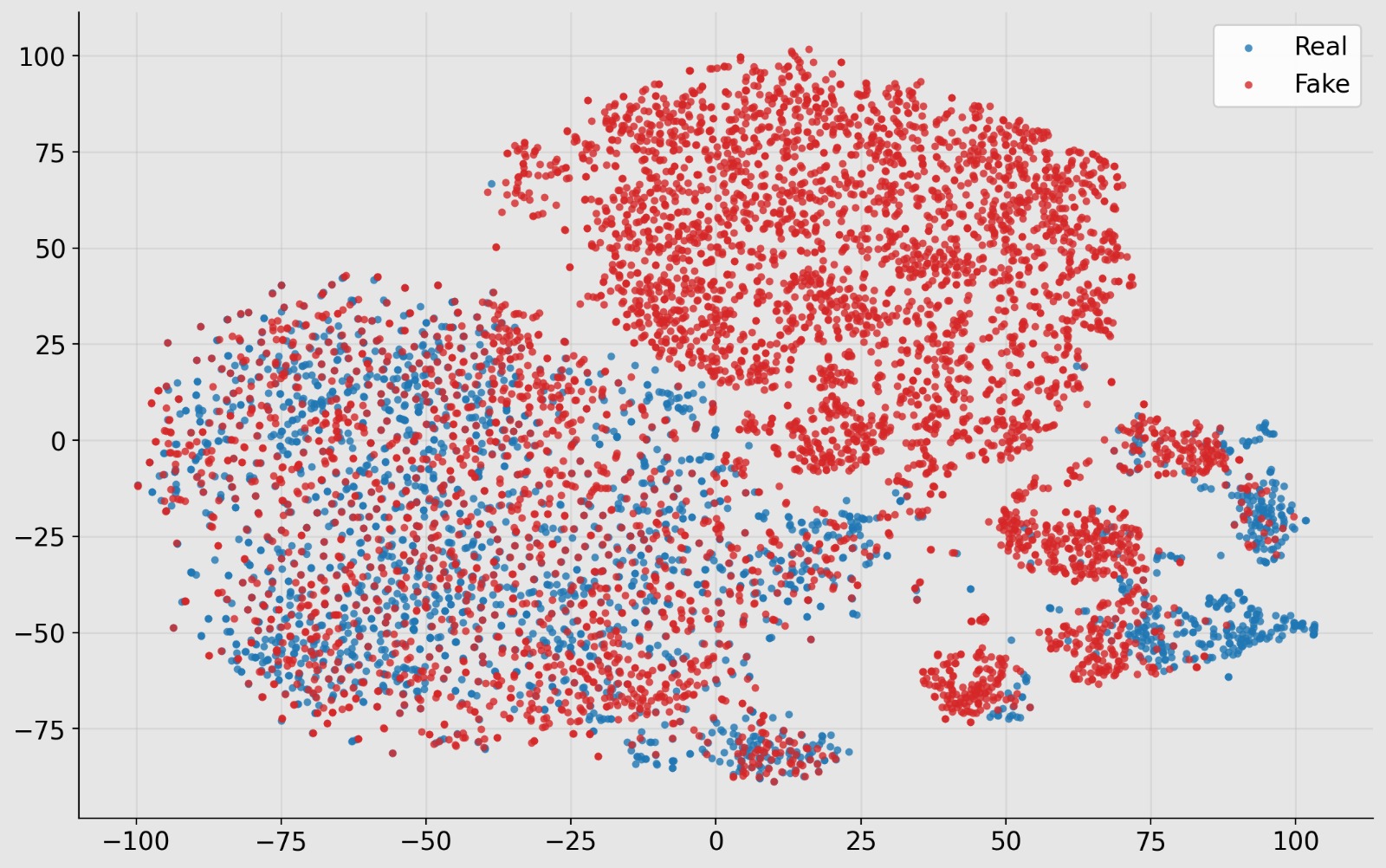}}
    \hspace{0.03\textwidth}
    \subfloat[]{\includegraphics[width=0.42\textwidth]{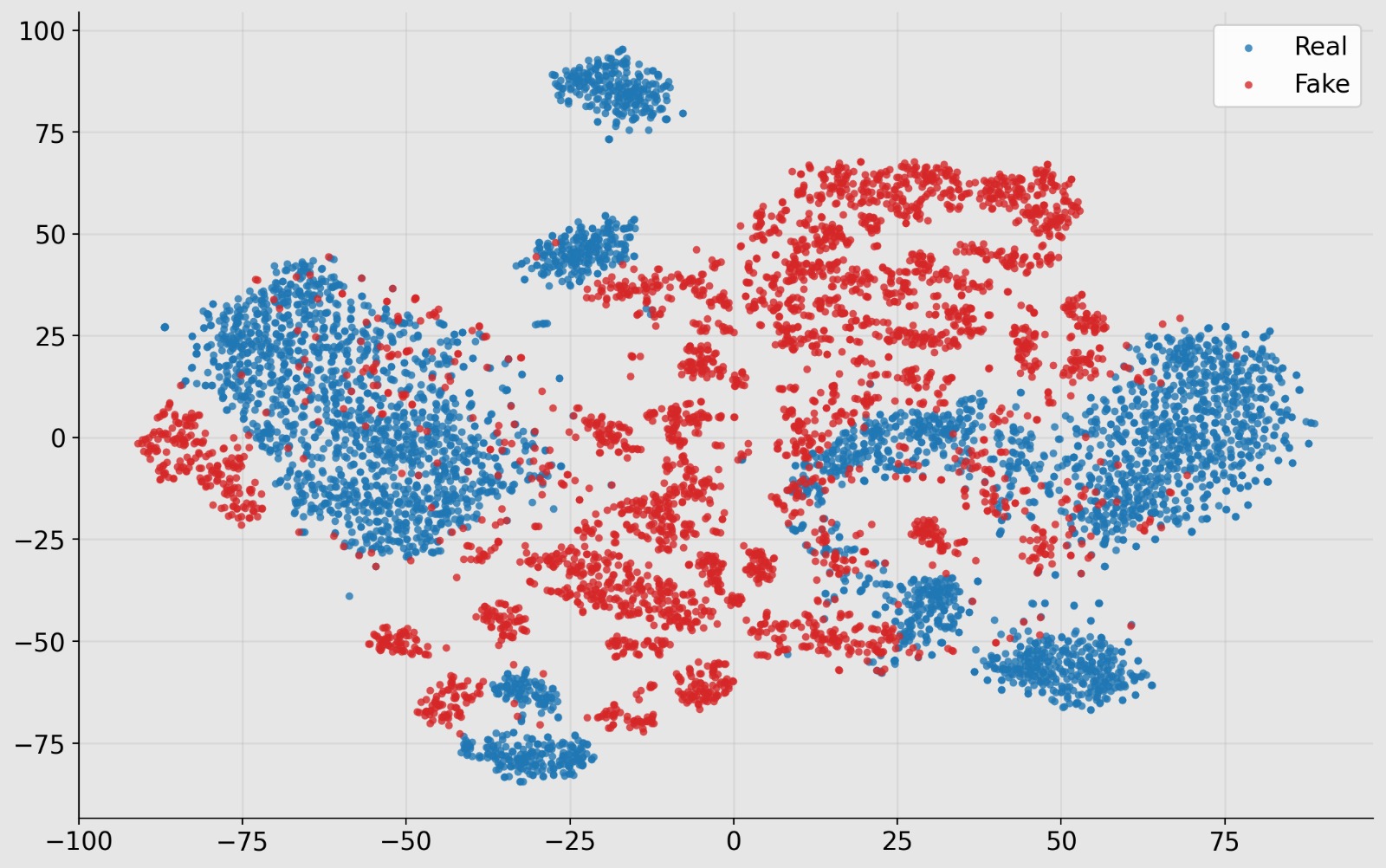}}\\[1ex]

    \subfloat[]{\includegraphics[width=0.42\textwidth]{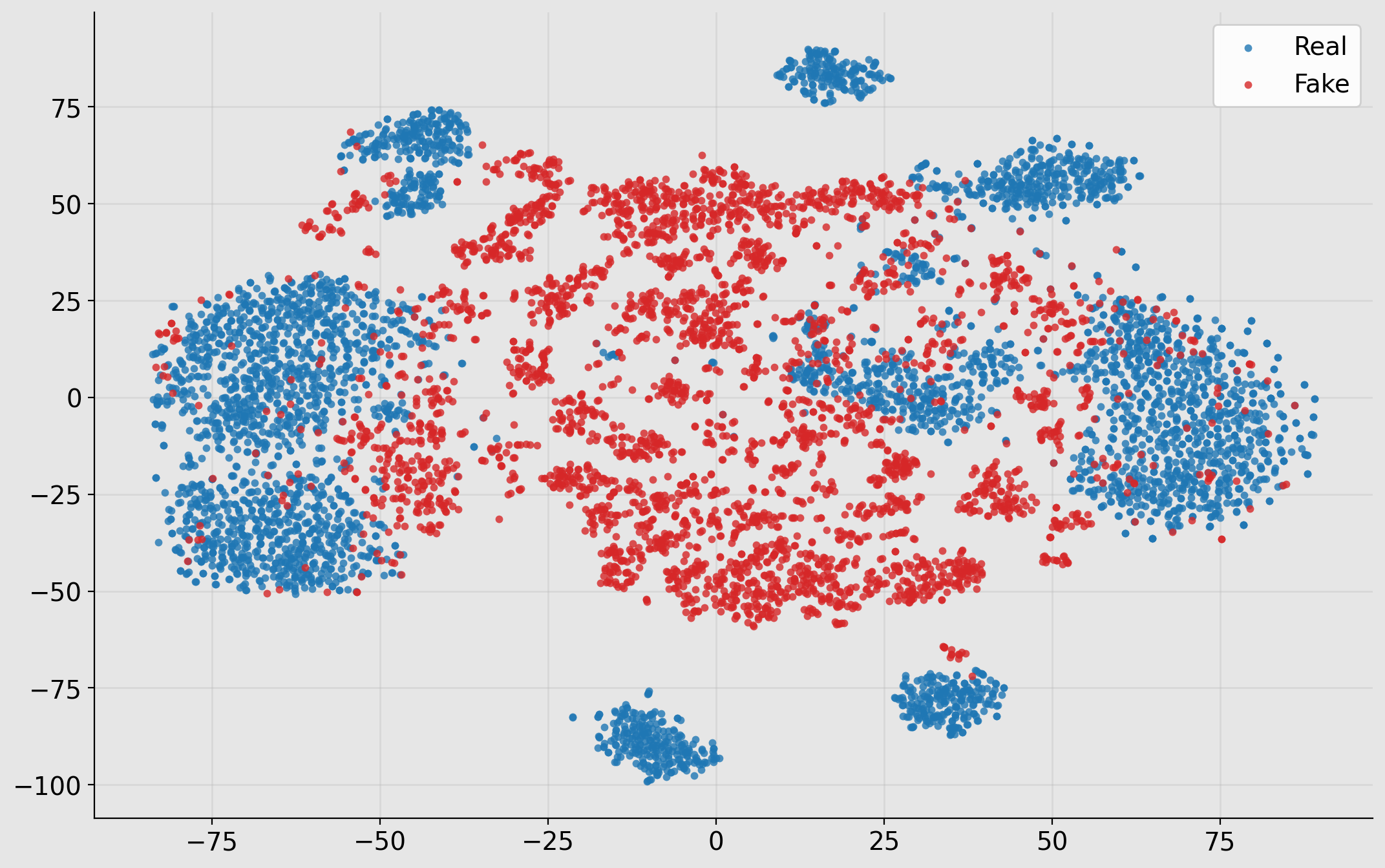}}
    \hspace{0.03\textwidth}
    \subfloat[]{\includegraphics[width=0.42\textwidth]{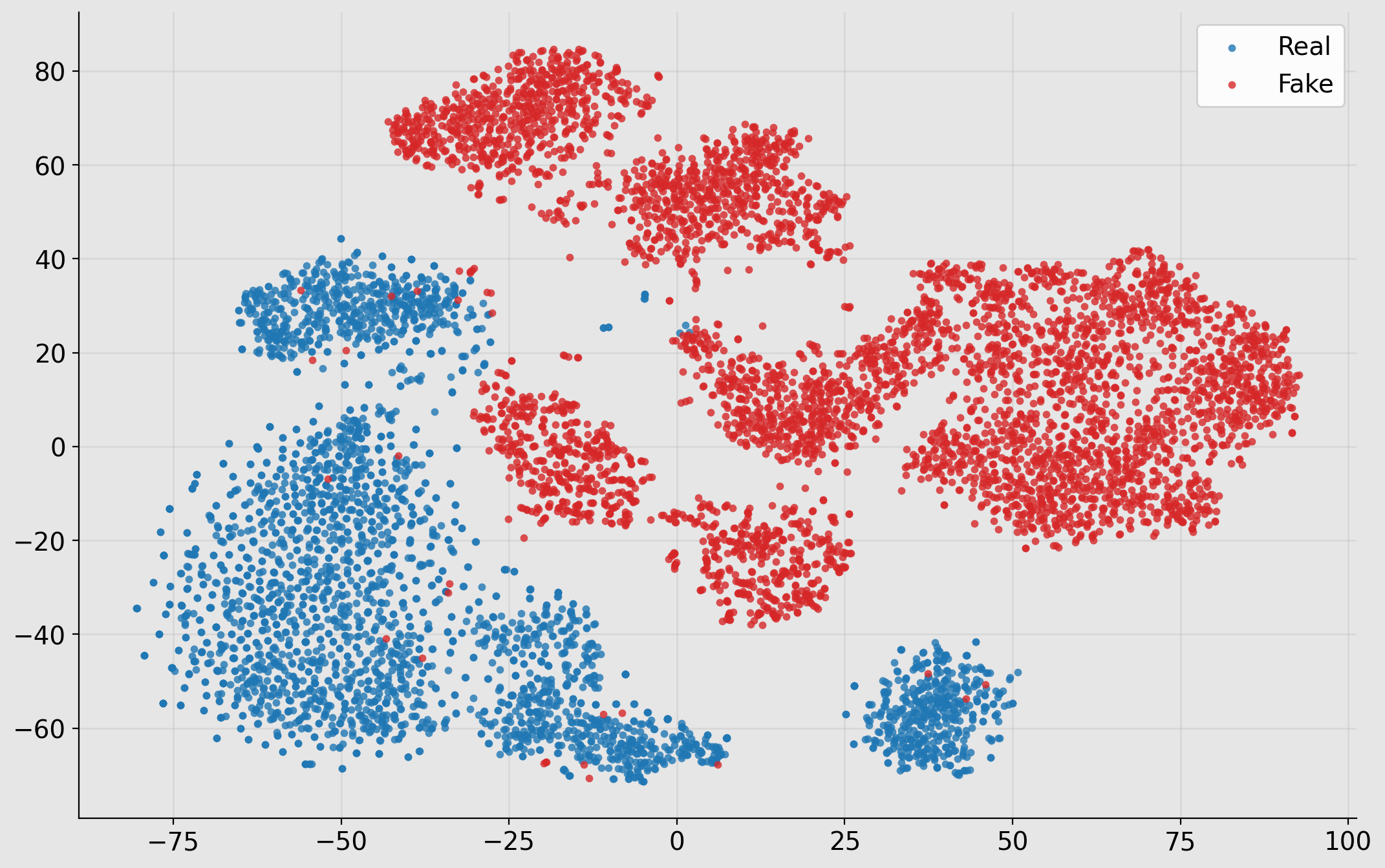}}\\[1ex]

    \caption{t-SNE visualizations: (a) Depression with PaSST on English \texttt{PHOENIX-Mamba}; (b) Alzheimer’s with WavLM on Chinese; (c) Alzheimer’s with WavLM on English; (d) Alzheimer’s with PaSST on English \texttt{PHOENIX-Mamba}; (e) Dysarthria with PaSST on English \texttt{PHOENIX-Mamba}; and (f) Dysarthria with PaSST on Chinese \texttt{PHOENIX-Mamba}. These plots provide an intuitive view of class separability.}
    \label{fig:tsne_8}
\end{figure*}

Figure~\ref{fig:tsne_8} visualizes learned utterance representations from selected \texttt{PHOENIX-Mamba} configurations. These projections provide an intuitive view of class separability and complement the class-wise error patterns observed in the main results.

\end{document}